\documentclass[%
amsmath,amssymb,
]{article}
\usepackage{authblk}
\usepackage{amsfonts}       
\usepackage{amsmath}
\usepackage{amssymb}
\usepackage{booktabs} 
\usepackage{graphicx}
\usepackage{dcolumn}
\usepackage{bm}
\usepackage[usenames,dvipsnames]{xcolor}
\usepackage[colorlinks=true,linkcolor=Blue,citecolor=Blue,urlcolor=Blue]{hyperref}
\usepackage[capitalise]{cleveref} 
\usepackage[a4paper, margin=3cm]{geometry}
\newcommand{\rev}[1]{\color{black} #1}
\newcommand{\revtwo}[1]{\color{black} #1}
\usepackage{tabularx}
\usepackage{lineno}

\begin{document}

\title{Evidence of social learning across symbolic cultural barriers in sperm whales}

\author[a,b]{Antonio Leitao}
\author[a,b]{Maxime Lucas} 
\author[a,b,k]{Simone Poetto} 
\author[c, d]{Taylor A. Hersh} 
\author[b,e,f]{Shane Gero} 
\author[b,g]{David F. Gruber} 
\author[b,h]{Michael Bronstein} 
\author[a,b,i,j]{Giovanni Petri}

\affil[a]{CENTAI Institute, Corso Inghilterra 3, 10138 TO, Italy}
\affil[b]{Project CETI, New York, NY, 10003 USA}
\affil[c]{Comparative Bioacoustics Group, Max Planck Institute for Psycholinguistics, Nijmegen, Netherlands}
\affil[d]{Marine Mammal Institute, Oregon State University, Newport, OR, United States}
\affil[e]{Department of Biology, Carleton University, Ottawa, Ontario, Canada}
\affil[f]{The Dominica Sperm Whale Project, Roseau, Dominica}
\affil[g]{Department of Natural Sciences, City University of New York, Baruch College, New York, NY, United States}
\affil[h]{Department of Computer Science, University of Oxford, Oxford, UK}
\affil[i]{NPLab, Network Science Institute, Northeastern University London, London, UK}
\affil[j]{IMT Lucca, Lucca, Italy}
\affil[k]{Center for Modern Interdisciplinary Technologies, Nicolaus Copernicus University, Toruń, Poland}

\date{\today}
\maketitle

\vspace{-2em}
\begin{abstract}
We provide quantitative evidence suggesting social learning in sperm whales across socio-cultural boundaries, using acoustic data from the Pacific and Atlantic Oceans. 
Traditionally, sperm whale populations are categorized into clans based on their \textit{vocal repertoire}: the rhythmically patterned click sequences (codas) that they use.  
Among these codas, \textit{identity} codas function as symbolic markers for each clan, accounting for 35-60\% of codas they produce. 
We introduce a computational method to model whale communication, which encodes rhythmic micro-variations within codas, capturing their \textit{vocal style}. 
We find that vocal style-clans closely align with repertoire-clans.
However, contrary to vocal repertoire, we show that sympatry increases vocal style similarity between clans for \textit{non}-identity codas, i.e. most codas, suggesting social learning across cultural boundaries. 
More broadly, this subcoda structure model offers a framework for comparing communication systems in other species,  with potential implications for deeper understanding of vocal and cultural transmission within animal societies. 
\end{abstract}

\section{Introduction}

Cultural transmission is broadly defined as the transmission of information or behaviors between individuals of the same species by means of social learning \cite{rendell_whitehead_2001, galef1992question}. 
While humans represent a benchmark of such capacity, cultural transmission has been observed in a wide variety of animals, including cetaceans \cite{whiten2017extension,whitehead2014cultural}, songbirds \cite{slater1986cultural}, non-human primates \cite{luncz2015primate}, and insects \cite{alem2016associative}. 
It typically takes one of three forms: vertical transmission, from adult kin to young kin; oblique transmission, from unrelated adults to young; or horizontal transmission, from peer to peer \cite{cavalli1981cultural}.

When animals have the capacity for social learning, group-specific differences can arise and remain stable when they become distinguishable by symbolic markers: arbitrary group-identity signals that are recognizable by both members of the group itself and by members of other groups~\cite{Barth1969,mcelreath2003shared}. 
In humans, symbolic markers can take a myriad of forms, ranging from visible signs, such as tattoos or garments, to communication cues or signals, such as idiomatic sentences or accents~\cite{Barth1969,mcelreath2003shared,BoydRicherson1987}. 
In animals, however, quantitative evidence of symbolic markers is remarkably scarce, one exception being recent results on the use of \textit{identity codas} in sperm whale social communication~\cite{hersh2022evidence}.  

\begin{figure*}[htb]
    \centering
    \includegraphics[width=0.9\textwidth]{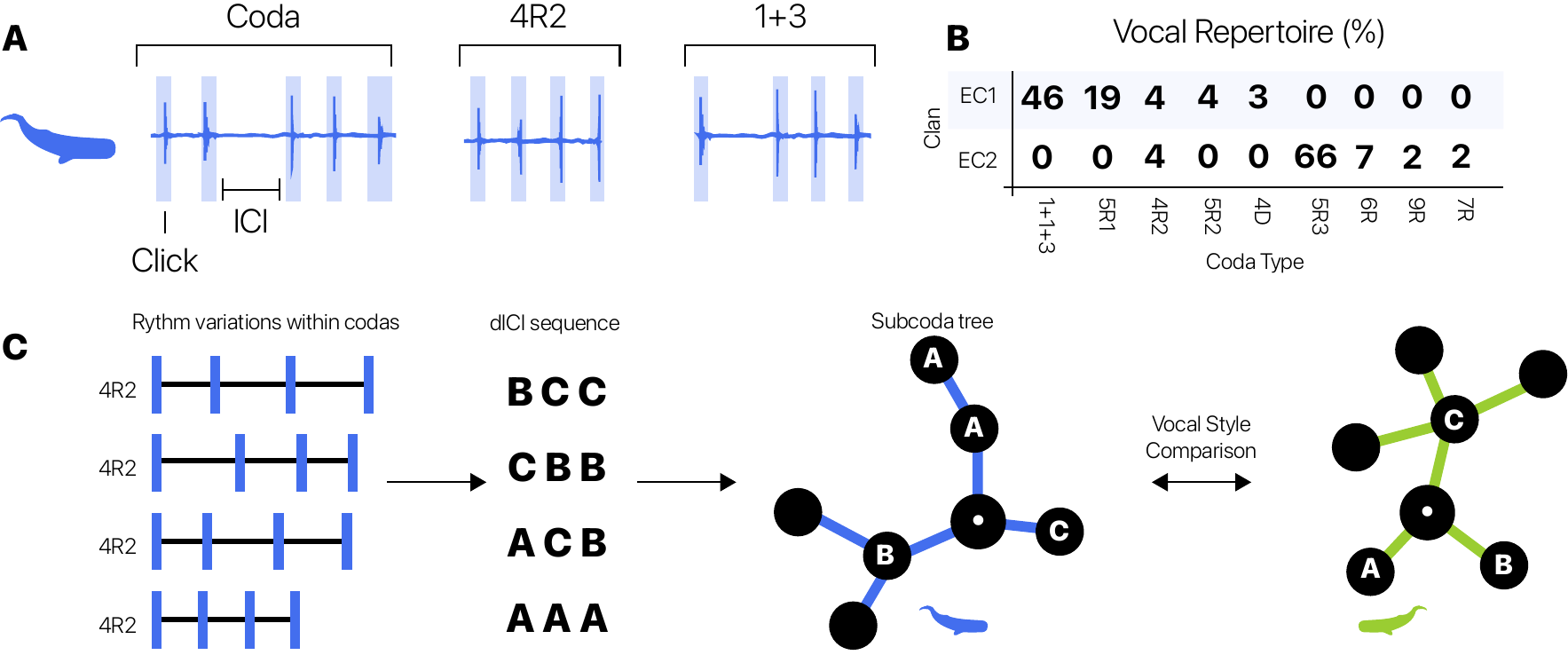}
    \caption{\textbf{Statistical modeling of subcoda structure in sperm whales. }  \textbf{A} Sperm whale communication consists of rhythmic sequences of clicks, called codas. 
    A coda is specified by a sequence of inter-click intervals (ICIs). 
    Codas are classified into types based on their rhythmic pattern, which can have various degrees of regularity (e.g., 4R2 vs 1+3). 
    \textbf{B} Social groups of sperm whales employ specific vocal repertoires: the set of coda types they use and their associated usage frequencies.
    As an illustration, we show those of the EC1 and EC2 clans from the Dominica dataset \cite{bermant2019deep}. 
    Only the most numerous coda types are shown: the rest of the vocal repertoires consists of more coda types with residual frequencies.
    \textbf{C}~The subcoda structure can be modeled by considering rhythmic variations within codas of the same coda type. 
    To do so, codas are represented as sequences of discrete inter-click intervals (dICIs), by discretizing absolute ICIs into discrete bins, which can then be considered akin to symbols (e.g. A, B, C,~...), providing a tokenization of codas. 
    Different instances of a single coda type can correspond to slightly different dICI sequences. 
    The resulting dICI sequences are modeled using variable-length Markov chains, which can be represented as \textit{subcoda trees}.
    {\rev These trees are built starting from the root node. Each node represents a dICI, and a dICI is a child of another in the tree if its presence in the affects the output probability of the parent, that is the probability of the parent depends on whether the child node has been observed previously in the sequence; see the Methods for a thorough explanation.}
    These trees can be built for an individual speaker or for a group of speakers {\rev (here represented by colors)}, and capture the statistical {\rev properties} of rhythm variations within codas and in the transitions between those. 
    In other words, the tree captures a vocal style---how they say what they say.    
    The vocal styles of different groups of sperm whales can be quantitatively compared by calculating a distance between their subcoda trees, {\rev which is a statistical distance between the dICI output distributions}.
    }
    
    \label{fig:figure_one}
\end{figure*}

Sperm whales live in multi-tiered societies and have a complex vocal communication system~\cite{SpermWhaleChapter}. 
They communicate through rhythmic patterns (\textit{codas}) of short  broadband sounds (\textit{clicks}), which have traditionally been classified into a set of \textit{coda types}. This classification is based on the total number of clicks and their rhythm and tempo extrapolated from the time interval between {\revtwo{the onsets of}} consecutive clicks: the inter-click interval (ICI) \cite{watkins1977sperm, hersh2021using} (\cref{fig:figure_one}A). 
{\revtwo{This measure is equivalent to the inter-onset intervals (IOIs) often used in rhythm analyses \cite{burchardt2020comparison,friberg2002structural,de2021born} but for the sake of compatibility with studies on sperm whale acoustics, we use ICI terminology throughout this paper.}}
For example, the 4-regular (4R2) type refers to a pattern of four evenly spaced clicks, whereas the 1+3 type refers to two clicks separated by a longer pause followed by two clicks in quick succession. 
Coda types are thus standardized rhythmic patterns, but individual vocalizations of a given coda type exhibit micro-variations around that pattern.
{\rev In the case of sperm whales ICIs can be used instead of IOIs due to the very low variance in duration of the individual clicks \cite{goold1995time,ravignani2019modelling}.}

The set of vocalized coda types (\textit{coda usage}) combined with how frequently each is vocalized (\textit{coda frequency}) makes up a \textit{vocal repertoire} (\cref{fig:figure_one}B). 
For example, the 4R2 coda is used by many sperm whales, but other coda types are more specific in their usage or frequency to certain groups of sperm whales.
While there is evidence of individual variation in vocal repertoires  \cite{weilgart1993coda,Schulz2010Individual,gero2015individual}, sperm whales belonging to the same social unit—a stable, matrilineally-based group of whales, share a common vocal repertoire that is stable across years \cite{Schulz2010Individual,gero2015individual,Rendell2005temporalstable}. 
Social units that share substantial parts of their repertoire are said to be part of the same \textit{vocal clan} \cite{rendell2003vocal,gero2016socially}. 
There is clear social segregation between members of different clans, even when living in sympatry, and thus clans mark a higher level of social organization, which appears to be defined on the basis of cultural vocal  markers~\cite{rendell2003vocal,gero2016socially,hersh2022evidence}
(see \cref{tab:concept_syummary} for a summary of the key concepts).

The clan specific and frequent usage of certain coda types, termed \textit{identity codas} \cite{hersh2021using, hersh2022evidence}, align with the expectations for symbolic markers of group membership \cite{mcelreath2003shared}. Furthermore, quantitative evidence that sperm whales themselves use identity codas as such markers has recently emerged: the more two clans overlap in geographic space, the more different their identity coda usage is \cite{hersh2022evidence}. This is consistent with computational models \cite{mcelreath2003shared} of the evolution of symbolic marking, which predict that differences between cultural norms will be starkest when inter-group interactions are more common (e.g., in boundary or overlap regions).  
  
All remaining coda types have been referred to as \textit{non-identity} (non-ID) \textit{codas} and constitute a very large fraction of sperm whales' total number of coda utterances. 
In fact, the total number of emitted non-ID codas accounts for more than 6 out of 10 codas (see SM Section 1.1 for the counts per clan and per coda type).
This begets the question: if ID codas are used as clan identity signals, what can be said about the remaining 65\% of codas?   

Here, we introduce a novel descriptive framework that focuses on the \textit{subcoda structure}, that is, the rhythmic micro-variations of intervals between clicks within codas (\cref{fig:figure_one}C). 
This framework, formally encoded in what we call a ``subcoda tree'', captures \textit{how} codas are uttered: a \textit{vocal style}. 
We find that variations in this vocal style, even for a single coda type, identify an individual's social unit and clan, effectively fingerprinting vocal repertoires. 
With this, we add a new dimension with respect to previous approaches based on \textit{which} codas are said---vocal repertoires. Thus, we propose a new concept of \textit{vocal identity} of sperm whales that comprises both vocal style and vocal repertoire. 


By applying our modeling framework to acoustic data from the Atlantic and Pacific  Oceans, we obtain two main results. First, we partition sperm whale populations into vocal style-defined clans, which we find to recapitulate the previously defined vocal repertoire clans. 
This confirms our method{\rev's validity as it reiterates previous studies' results}. 
Second, and crucially, we find that the vocal style of non-ID codas is more similar for more sympatric clans, i.e. clans whose territory overlaps more spatially. 
In contrast, we do not find an effect of sympatry on the similarity of vocal styles when studying only ID codas.  
This suggests that geographic overlap induces vocal styles to become more similar between clans, without jeopardizing each clan's acoustic identity signals. 
Our results strengthen previous results on the use of ID codas as symbolic markers, while 
supporting cultural transmission and social learning of vocalizations among whales of \emph{different} clans, as predicted by theoretical models \cite{cantor2015multilevel}.

\begin{table}[h]
    \centering
    \begin{tabularx}{\textwidth}{lX}
        \toprule
        \textbf{Concept} & \textbf{Description}\\
        \midrule
        Click & Sharp broadband pulses \cite{backus1966physeter} \\
        Inter-click interval (ICI) & Time interval between two consecutive clicks \\
        {\rev Discrete ICI (dICI)} & Discretization of an inter-click interval given a bin size. \\
        Coda & Stereotyped, short series of clicks produced by sperm whales in social situations \cite{watkins1977sperm}\\
        Coda type & Categorical classification of codas based on rhythm and tempo of coda inter-click intervals \cite{weilgart1993coda} \\
        Coda frequency & Usage frequency of each coda type\\
        Coda sample & Set of codas recorded during one event \\
        Coda usage & Set of coda types used by a group of whales \\
        Identity codas & Subset of coda types most frequently used by a sperm whale clan; canonically used to define vocal clans. \\
        Non-identity codas & Coda types produced by whales belonging to multiple vocal clans (i.e., all coda types that are not identity codas) \\
        Vocal clan & Higher-order level of social organization delineated based on social units that share a substantial part of their vocal repertoire \cite{hersh2022evidence,rendell2003vocal}\\
        Vocal repertoire & Coda usage and frequency combined. Description of what sperm whales vocalize. \\
        Vocal style & Encodes the rhythmic variations within codas. Description of how sperm whales emit codas \\
        Vocal identity & Concept comprising both vocal style and vocal repertoire \\
        \bottomrule
    \end{tabularx}
    \caption{Summary of key concepts}
    \label{tab:concept_syummary}
\end{table}

\section*{Results} \label{sec:results}

\subsection*{Subcoda structure captures variability in sperm whale communication} 

\begin{figure*}[hbt]
\centering
    \includegraphics[width=0.9\textwidth]{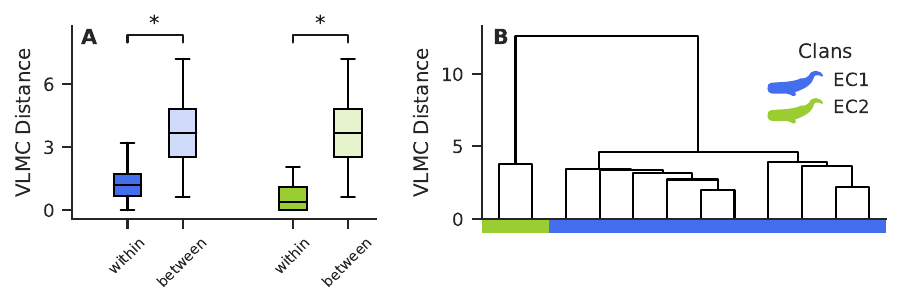}
    \caption{\textbf{Vocal style recovers social structure of vocal clans in Dominica sperm whales.} 
     \textbf{A} We show the similarity of vocal style, measured as subcoda tree-distance, among social units within a vocal clan (\textit{within}, darker color shade) and between two clans (\textit{between}, lighter color shade). We used the manual clan assignments from \cite{gero2016socially} as ground truth. Vocal style is more similar within clans than between clans.
     \textbf{B} We show the hierarchical clustering of social unit subcoda trees. Each leaf corresponds to a social unit, and the colors below show their known clan assignments. The clustering recovers the two-clan structure observed in past work \cite{gero2016socially}. }
     \label{fig-validation-dominica}
\end{figure*}

We model the internal structure of codas, in terms of rhythmic variations at the level of clicks, by using variable length Markov chains (VLMCs). 
Our analytical pipeline is illustrated in \cref{fig:figure_one}C. 
We build each VLMC in two main steps. 
We first convert codas, naturally represented as sequences of continuous, absolute, inter-click intervals (ICIs), to sequences of discrete ICIs (dICIs), by discretizing time into bins. 
In this way, each dICI represents a narrow range of possible ICI values. 
The bins have a fixed width (or resolution) $\delta t$ and thus implicitly correspond to the temporal resolution of our representation (see Methods for details on the optimal choice of $\delta t$). 
Note that although ICIs have units of time (seconds), dICIs are (unit-less) symbols (e.g. A, B, C, etc.), representing multiples of $\delta t$ (and so the smaller $\delta t$, the more the symbols).
For example, the shortest ICIs will be mapped to the symbol A whereas longer ones will be mapped to symbols further down the alphabet.
Hence, each coda (a sequence of ICIs) is mapped to a sequence of discrete symbols (a sequence of dICIs).  
The second part of the pipeline focuses on modeling the internal structure of codas in terms of dICI sequences.
Essentially, we want to estimate transition probabilities from a dICI sub-sequence to the next dICI.  
A standard way would be to describe this using $k$-order Markov chain models, {\rev which model the next state (dICI) probability given a past sequence of $k$-states (dICIs).} 
However, it is possible that specific {\rev dICIs depend on sequences of varying length}. 
{\rev Despite we do not have direct evidence of unitary blocks in sperm whale communication, on can imagine this effect similarly to what happens with words (e.g., a word beginning with ``re" can continue in more ways than one starting with ``zy").} 
To account for {\rev the possibility that states (words, dICIs) might depend on sequences of varying length,} while also retaining only the most compressed statistical representation of how 
codas are structured in terms of dICIs, we employ VLMCs. 

VLMCs are generalizations of standard (fixed-memory) Markov chains that allow sub-sequences of dICIs of variable lengths. Longer sequences are kept only if they are significantly more informative in predicting the next dICI than random chance, yielding an optimally compressed representation
(see Methods for details on model fitting and selection, including the optimal choice of $\delta t$). 
Furthermore, VLMCs naturally have a tree structure (see \cref{fig:figure_one}C),  because of the natural order between sequences and their sub-sequences.
In particular, each node represents a sub-sequence of dICIs, and is equipped with a probability distribution of transitions to the next dICIs.
The origin node corresponds to the empty sequence, leaf nodes correspond to the longest sequences, and all nodes forming the branch in between correspond to the sub-sequences of that leaf node. 
Thus, we call VLMCs fitted to coda ICI data \textit{subcoda trees}.

Note that dICI sequences encode rhythmic variations within codas. Indeed,  a coda type is a standard rhythmic pattern that can be realized with variations in its ICIs and thus in its dICIs too. For example, the 4R2 coda type can be vocalized as BCC but also as CBB (in a representation with, say, 26 symbols). In that sense, subcoda trees, through the dICIs sequences that they contain and their transition probabilities, capture information about a vocal style.
Four more features of subcoda trees are noteworthy: 
(i) because the method's input is a set of codas, we can build subcoda trees for repertoires corresponding to different social scales, from individual sperm whales, to social units, all the way up to vocal clans;
(ii) the difference between different subcoda trees can be measured using a probabilistic distance (see Methods), which we can use to compare subcoda trees across sperm whale clans;
(iii) certain features of the vocal style can be quantified via metrics on the subcoda tree: for example we can define a complexity of the vocal style measured by an entropy on the tree;
and (iv) subcoda trees can also be used as generative models, to create new synthetic codas in the form of dICI sequences to train downstream machine learning models. 

\subsection*{Vocal style recovers vocal clan structure}
\begin{figure*}[htb]
\centering
    \includegraphics[width=\textwidth]{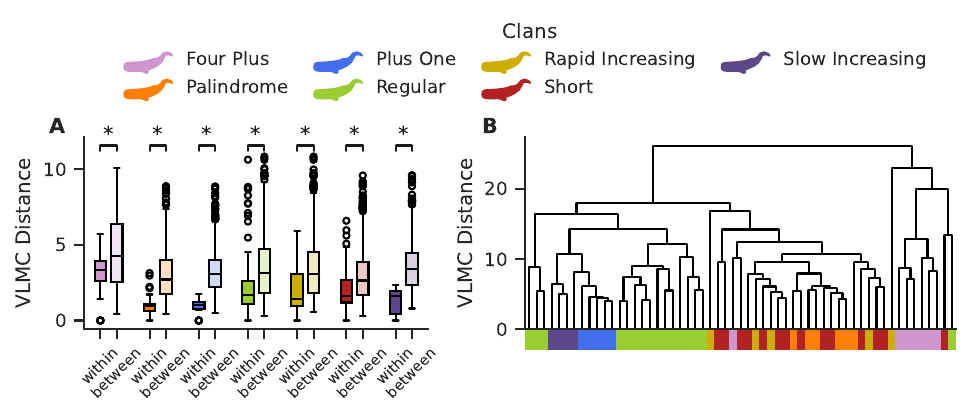}
    \caption{\textbf{Vocal style recovers social structure of vocal clans in Pacific Ocean sperm whales.} 
     \textbf{A} We show the similarity of vocal style, measured as subcoda tree-distance, between coda samples within a vocal clan (\textit{within}, darker color shade) and between a clan and all others (\textit{between}, lighter color shade). We used the vocal clans identified in \cite{hersh2022evidence}. Vocal style is more similar within a clan than between clans.
     \textbf{B} We show the hierarchical clustering of subcoda trees. Each leaf corresponds to a coda sample, and the colors below show their vocal clan assignments (based on coda usage) from \cite{hersh2022evidence}. 
     We find generally good overlap between the groups obtained from clustering vocal style and those from vocal repertoire, with the exception of the \textit{Short} clan (red) that is somewhat mixed with the \textit{Palindrome} (orange) and \textit{Rapid Increasing} (yellow) clans.}
    
     \label{fig-results-pacific}
\end{figure*}
The information about vocal style contained in subcoda trees is sufficient to recover the social structure of sperm whales (social units and clans). 
We show this in two ways. 
First, we analyze a dataset from sperm whales in Dominica (Dominica dataset) \cite{gero2016socially}. 
This dataset has rich annotations (coda type annotations, identity of recorded whales, social relations of recorded whales) which makes it particularly useful for validation. 
Specifically, the sperm whales in the Dominica dataset are divided into well known social clans, each composed of several social units,each with its own specific vocal repertoires, and thus can be defined as two different vocal clans. 
For each social unit in this dataset, we aggregate the individual whales' coda samples and build a subcoda tree. 
Computing the distance between these trees (see Methods), we find that the distances between social units within the same clan are significantly smaller than between clans (\cref{fig-validation-dominica}A).  
We also find that an agglomerative clustering (average linkage, see Methods for details) on the distance between the subcoda trees correctly clusters social units into their respective clans (\cref{fig-validation-dominica}B). 
Without a priori knowledge of the clan memberships, we used vocal style to recover the existing classification of social units into two clans, which was previously done based on similarity between vocal repertoires (i.e., coda types and usage) \cite{gero2016socially}.

Second, we find that the subcoda structure of synthetic codas, generated from subcoda trees fitted on real data, closely reproduces that of real codas. 
To do this, we first train a simple classifier to assign codas to one of the two vocal clans, based on coda type. 
Variations of the same classifier, trained on the same real data, have been shown to discriminate between individual whales, social units, and clans with high accuracy \cite{bermant2019deep}. 
We train the classifier on real codas, and then test it on both real and synthetic ones.
The synthetic codas were generated using the subcoda tree of each clan, with a number of codas similar to that of the original dataset for a fair comparison (see Methods for details).
We find that synthetic codas are correctly classified into their clans with an accuracy close ($\sim85\%$) to that obtained on the real data ($\sim90\%$, see \textit{Supplementary Materials} Section $4$). 

Motivated by these results, we extend our analysis to a much larger dataset from the Pacific Ocean (Pacific dataset) \cite{hersh2022evidence}. 
This dataset is more sparsely annotated because of the breadth of its spatial coverage.
We restricted our analyses to a well-sampled subset ($n = 57$ coda samples) of the full Pacific dataset (see Methods for details).
Coda samples are only labeled by the spatial position at which they were recorded, but no information is available about the identity of the vocalizing sperm whales (see Methods for details). 
In fact, each repertoire likely contains codas from multiple individuals of a single clan.
It has recently been shown that these coda samples can be divided into seven vocal clans based on their coda usage \cite{hersh2022evidence}. 
We use those clans as a benchmark for the following analysis. 

Since there is no social unit-level information for this dataset, we fit a subcoda tree for each repertoire (i.e., all of the codas recorded on a single day in a single region). 
Trees are significantly more similar for coda samples belonging to the same vocal clan than for those belonging to different vocal clans (\cref{fig-results-pacific}A). 
We also find that clustering coda samples based on vocal style returns a dendrogram that closely matches the one obtained from coda usage in \cite{hersh2022evidence} (\cref{fig-results-pacific}B). 
The major exception we find is the \textit{Short} clan (red), named because member whales produce short codas with very few clicks, for which anomalous results were previously reported as well \cite{hersh2022evidence}.
In our case, this is due to the Short clan being less well localized in the space of trees, while the other clans have well-defined centroids (see \textit{Supplementary Materials} Fig. $9$ for a low-dimensional representation subcoda tree metric space). 

Therefore, we find that sperm whale vocal clans in the Atlantic Ocean (Caribbean Sea) and Pacific Ocean can be identified by a \textit{vocal identity} that encompasses both clan-specific \textit{ vocal repertoire} \cite{gero2016socially, Vachon3clans, rendell2003vocal, hersh2022evidence} and \textit{vocal style} as defined in this work.

\subsection*{Clan sympatry impacts vocal style of non-ID codas only}
\begin{figure*}[htb]
    \centering
    \includegraphics[width=\linewidth]{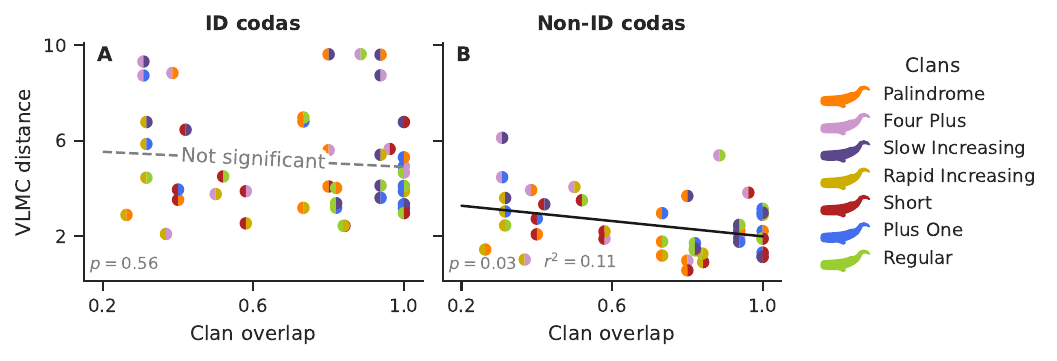}
    \caption{\textbf{Clan overlap influences non-ID coda vocal styles} Comparing the similarities of different VLMC models fit for each Pacific Ocean clan for both ID and non-ID coda samples. The y-axis represents the measured distance between the subcoda trees, and the x-axis shows the geographical clan overlap (as calculated in \cite{hersh2022evidence}{\rev; see "Measuring clan overlap" in our Methods for details}). Each point represents a pairwise comparison between two clans. The effect of overlap on ID coda vocal style similarity is minimal and non-significant while the opposite is true for non-ID codas: overlapping clans produce non-ID codas with a more similar vocal style. The VLMC distances are also typically much greater for ID codas than for non-ID codas.
    Note that while these results are visually opposite to those reported in Hersh et al. \cite{hersh2022evidence}, they support the same final conclusions (see ``Identity and non-identity codas show different trends" for details).}
    \label{fig:similarity-clan-overlap}
\end{figure*}
While interesting, the fact that both vocal repertoires and vocal styles discriminate between clans might imply that considering both could be redundant for vocal identity. 
However, we find that this is not the case when we consider the functional role of ID versus non-ID codas. 

More precisely, different clans can share significant portions of their total range, overlapping across large swaths of ocean. 
Such sympatric clans exhibit a decreasing similarity of their ID coda usage with increasing clan overlap \cite{hersh2022evidence}. 
This means that the more two clans overlap in space, the more dissimilar their vocal repertoires are in terms of ID coda types and their usage frequency. 
This is consistent with the idea that ID codas are used as symbolic markers to delineate cultural boundaries between social groups \cite{hersh2022evidence, mcelreath2003shared}. 
In contrast, non-ID coda usage do not show any relationship to clan overlap.  

We find the exact \textit{opposite} effect when considering vocal style. 
The similarity in vocal style for ID codas across clans does not depend on the level of clan overlap (\cref{fig:similarity-clan-overlap}a). 
In contrast, the similarity in vocal style for non-ID codas displays a clear and significant increase (i.e., decreasing subcoda tree-distance) as clan spatial overlap increases (\cref{fig:similarity-clan-overlap}b). 
In the \textit{Supplementary Materials} (see Section $2.4.2$), we show that the same results hold at the single coda type level, in addition to the whole clan level, along with an analysis of the confidence intervals.
These results imply that the internal structure of codas
is more similar for groups that likely spend more time in the same space, {\rev which can be thought to be analogous to} accents aligning in human populations that share the same territory~\cite{cohen2012evolution,henrich2021origins}. 
This also highlights the complementarity of vocal repertoire and style: the trends are different precisely because the two concepts describe different aspects of whale vocalization.

\section*{Discussion}
We have presented a general method for modeling animal communication systems and their complexity based on VLMCs. 
In the context of sperm whales, this new method allows the extraction of \textit{subcoda trees}, which succinctly describe the internal temporal structure of codas. 
Previous work on the structure of sperm whale communication has largely focused on supra-level coda analyses: for example, by classifying codas into types, quantifying how often different types are used, and distinguishing between individual whales, social units, or clans based on those counts \cite{gero2015individual, oliveira2016sperm}. 
Here, we adopted a more fine-scale approach by investigating potential structure \textit{within} codas. 
To do so, we used VLMCs to model the transition probability of observing a specific ICI given the previous ones. 
A VLMC, or here a subcoda tree,  encodes all those probabilities but only for dICI sequences that are informative---other sequences are automatically discarded. 
As such, a subcoda tree is a statistically validated representation of the internal memory structure of codas at the level of sequences of clicks. 
It contains information about important rhythmic variations and transitions between them: a vocal style.

Using such representations, we propose a novel concept of \textit{vocal identity} for sperm whales composed of \textit{vocal repertoire} (what they say) and \textit{vocal style} (how they say it), the latter being captured by our framework. 
We find that: 
(i) vocal styles vary between social units and clans, and can be used to distinguish them;  
(ii) the similarity of clan vocal styles for non-ID codas increases with increasing spatial overlap, while no change occurs for ID codas; and
(iii) social learning across symbolic cultural boundaries most parsimoniously explains the observed trends.

\subsection*{Vocal style recovers hierarchical social structure}
Using the Dominica dataset, sperm whales had previously been divided into two vocal clans, based on their vocal repertoires and observed social interactions \cite{gero2016socially}.
In our study, comparing the vocal styles of those same whales led to the same assignment of social units to two vocal clans. 
Similarly, for the Pacific dataset, clustering based on vocal styles yielded clans that were in good agreement with those previously defined based on vocal repertoires \cite{hersh2022evidence} (\textit{Supplementary Materials} for an extended comparison).
The difference between the two partitions was mainly due to the Short clan, which was more spread out in subcoda tree space than the other clans, causing overlap with other clans that showed less variability. 
This variability could be linked to the fact that Short clan whales typically make codas with very few (e.g., three or four) clicks, leading to subcoda trees with very few nodes. 
In Ref. \cite{hersh2022evidence}, the authors observe a similar lack of uniformity in coda usage of the Short clan.

\subsection*{Identity and non-identity codas show different trends}

For ID codas, we show that the similarity between clan vocal styles is not affected by spatial overlap, while it has recently been shown that the similarity between clan vocal repertoires decreases with overlap {\rev (measured on the two clans' joint set of ID codas) \cite{hersh2022evidence}}. 
This means that spatial overlap does not affect \textit{how} whales produce ID codas (in terms of their fine-scale rhythmic structure; our results) but does affect \textit{how often} they produce them {\rev \cite{hersh2022evidence}}.
In contrast, for non-ID codas, we show that the similarity between vocal styles increases with spatial overlap between two clans, while no change was observed for vocal repertoires in previous work on the same dataset.
In other words, increasing spatial overlap is correlated with more similar fine-scale rhythmic structure of non-ID codas produced by whales from different clans (our results), but does not affect how often non-ID codas are produced.
Our study thus supports and nuances the results of Hersh et al. \cite{hersh2022evidence}.
We provide further support for 
selection acting to produce unambiguous, recognizable identity signals in the ID codas. 
However, ID codas only account for 35\% of the total vocalizations; the remaining 65\% of codas have traditionally been lumped into a catch-all category (i.e., non-ID codas) and their function remains enigmatic (these numbers are an average over the Pacific clans, and go up to 93\% for non-ID codas when counting number of coda types instead of number of codas emitted, see SM 1.1 for details).
We could still discriminate among clans using non-ID coda vocal style; however, the 
increased similarity of non-ID coda vocal styles between clans with greater spatial overlap, as
demonstrated here, suggests that non-ID codas are likely vocal cues and not identity signals like the ID codas. 
Accordingly, vocal repertoire and vocal style capture different and complementary information on sperm whale communication, and should be considered in tandem in future studies.

\subsection*{Evidence for social learning across cultural boundaries}
There are several potential mechanisms driving the similarity in non-ID coda vocal styles---but not ID coda vocal styles---across spatially overlapped clans: environmental variation, genetics, and/or social learning.

Local adaptation to specific ecological conditions can lead to geographic variation in acoustic signals \cite{sun2013geographic}. 
If environmental pressures alone were responsible for the trends we observe in sperm whales, this would imply that (i) more spatially overlapped clans experience more similar environments, (ii) non-ID coda vocal style is impacted by or dependent on environmental parameters, and (iii) ID coda vocal style is not impacted by/dependent on environmental parameters.
Although the first point is somewhat intuitive, to date there is no evidence that coda production systematically varies with environment.
{\rev In fact, clans are recognizable across ocean basins, making local adaptation an unlikely driver of the observed trend in non-ID coda vocal style.}

If genetic relatedness were responsible, this would imply that (i) more spatially overlapped clans are more genetically related, (ii) non-ID coda vocal styles are genetically inherited, and (iii) ID coda vocal styles are not genetically inherited. If all three requirements were met, then the observed similarity in non-ID coda vocal styles for more spatially overlapped clans could be due to genetic determination under a general isolation by distance structure. However, research to date suggests this scenario is unlikely. Rendell et al. \cite{rendell2012clangenetics} found little evidence to support genetics as an explanation of differences in vocal dialects among clans in the Pacific Ocean. Furthermore, Alexander et al. \cite{alexander2016influences} found that regional genetic differentiation in the Pacific Ocean is very low: while social group is important for explaining both mitochondrial and nuclear DNA variance, geographic region is not. This contrasts with results from the Indian Ocean, where region was the strongest predictor of mitochondrial DNA variance. Given that gene flow in sperm whales is largely male-mediated and that mitochondrial DNA haplotypes are broadly shared across the Pacific Ocean, it is unlikely that coda dialects are genetically determined \cite{alexander2016influences,lyrholm1999sex}. Agent-based models grounded in empirical data from Pacific Ocean sperm whales further support {\rev coda usage} as socially learned, not genetically inherited \cite{cantor2015multilevel}. To fully rule out a genetic explanation for our results, the analyses in \cite{rendell2012clangenetics} could be replicated for ID coda {\rev usage} and non-ID coda {\rev usage} separately. This would shed light on whether certain coda types are genetically inherited vs. socially learned, as has been suggested for some humpback whale (\textit{Megaptera novaeangliae}) vocalizations \cite{epp2021allopatric}.


The most parsimonious explanation for the observed similarity of non-ID coda vocal styles of clans with increasing spatial overlap is social learning across clan boundaries. 
This is remarkable, given that sperm whales belonging to different clans have rarely been observed physically interacting at sea \cite{whitehead2014cultural}. However, that does not preclude the possibility that they are within acoustic range of each other \cite{madsen2002male} and that cross-cultural social learning opportunities arise. 
This explanation is compatible with (and bolsters) past work suggesting that ID and non-ID codas function differently in sperm whale communication, and further suggests that they experience different evolutionary pressures \cite{cantor2015multilevel}. Whether social learning has facilitated stochastic (i.e., cultural drift) or deterministic (i.e., cultural selection) processes is more difficult to determine, and it is unclear whether the observed non-ID coda vocal style alignment has been neutral or adaptive \cite{deecke2000dialect,sun2013geographic}. 
Importantly, these findings suggest that vocal learning in sperm whales may not be limited to vertical transmission from related adults to young kin, but that horizontal and/or oblique social learning from outside the natal social unit might also be occuring. 

Vocal identity in sperm whales is thus consistent with both cultural selection on ID codas to maintain discrete signals for vocal recognition in sympatry, and social learning between clans leading to a vocal style more similar to that of other whales with which they are in acoustic contact more frequently. This highlights a more complex system of transmission in which clan identity is maintained through selection, while gradual change over time may occur within and across clans for vocalizations which do not function in social recognition and thus may create similar vocal styles. 
{\rev Indeed, the observed alignment of non-ID vocal styles may reflect a combination of automatic imitation, and controlled imitation, guided by social and ecological factors, as discussed in studies of social learning mechanisms \cite{heyes2011automatic,heyes2012whats}. 
In the context of our findings, the similarity in non-identity codas across clans with geographic overlap may indicate an automatic component of social learning. Whales sharing the same habitat are exposed to overlapping acoustic environments, which could facilitate the development of similar vocal styles described here.
However, the persistence of distinct identity codas within clans may suggests a more directed form of social learning, shaped by social goals such as clan recognition and cohesion.}

\subsection*{Future directions}
Our results can be expanded in multiple ways in future work. 
The first, and the simplest conceptually, would be to conduct the present analysis on a larger dataset. 
More codas would improve the quality of the statistical analyses and ensure that all codas are represented in realistic proportions for each clan. 
Moreover, longitudinal datasets might provide direct evidence to discriminate between the social learning hypothesis and competing ones (e.g. drift in vocal style).
Similarly, confirmations could emerge from large scale genetic datasets addressing the issues of phylogenetic relatedness (or lack thereof) in clans that are closer in vocal style distance.
Such datasets do not exist at present, but efforts towards automated and semi-automated collection techniques are underway (e.g. Project CETI \cite{andreas2021cetacean}). 
Second, from a methodological perspective, we could add spectral information (in terms of acoustic frequencies) to the temporal information currently used. 
Although sperm whale acoustic communication seems mostly based on rhythm, spectral features of individual clicks may convey additional information. 
This possibility could be incorporated into our method by labeling the dICIs according to the frequency content of the associated click (or by extending the available ``alphabet" for the VLMC). 
Third, it would be interesting to investigate in more detail the function of non-ID codas. Indeed, even though ID codas were only recently formally named for the first time, they have been the primary focus of sperm whale coda research for decades. As previously mentioned, non-ID codas are a catch-all category for anything that is not an ID coda, but that does not mean that all non-ID codas function in the same way. To start to unveil  their function, we need to consider the context (behavioral, environmental, etc.) in which different non-ID codas are produced \cite{frantzis2008male}. The pattern we documented may or may not apply to all non-ID codas, but it is at least strong enough that we detect the relationship with clan spatial overlap when collectively considering all non-ID codas.
{\rev Finally, the methods we developed here could in principle be used to capture the structural complexity of any communication system, including those coming from other animal species, provided they can be encoded in sequences of tokens (e.g. calls, signs, etc.). 
This would also allow to compare the complexity of communication systems across species, possibly linking it to other ecological parameters of the environments inhabited by such species.}

\section*{Methods}
\label{sec:methods}
\subsection*{Acoustic data}

We analyzed two datasets in the present study. The Dominica dataset contains 8719 annotated codas recorded in the Atlantic Ocean off the island of Dominica between 2005 and 2019.
The codas come from 12 social units grouped into two vocal clans (EC1 and EC2). The Pacific dataset was collected between 1978 and 2017 at 23 locations in the Pacific Ocean (the recording methods are available in the supplementary materials of \cite{hersh2022evidence}). The codas were divided into coda samples according to their recording day and each repertoire was assigned a single vocal clan inferred in \cite{hersh2022evidence}. When considering a clan-level analysis (\cref{fig-results-pacific}) all coda samples were used to compute the subcoda trees (23555 codas). However, when analysing at a coda samples level (\cref{fig:similarity-clan-overlap}), we discarded coda samples with less than 200 codas with statistical inference in mind, resulting in a final count of 57 coda samples (17046 codas) for the Pacific.

\subsection*{Representation of sperm whale communication as discrete inter-click intervals}
As a preliminary step, we discretized the (continuous) ICI values into bins of width $\delta t$ seconds. In other words, we represented the continuum of ICI values by a finite set of discrete ICIs (dICIs) based on the duration of the ICI. The bin width $\delta t$ controls the \textit{temporal resolution} of the representation: a higher value of $\delta t$ implies a coarser representation with fewer dICIs.
We also imposed an upper bound $t_{\text{max}}$: any ICI value greater than that was truncated to $t_{\text{max}}$. This ensured that the set of dICIs was finite. Note that although ICIs have units of time (seconds), dICIs are unitless (they represent time intervals).
The resulting representation of ICIs as dICIs is a discrete random variable defined as
%
\begin{equation}
X_{\delta t,t_{\text{max}}} = \left\lfloor\frac{\min\left(\text{ICI}, t_{\text{max}} \right)}{\delta t}\right\rfloor ,
\end{equation}
which takes values in the finite set $\mathcal{X}=\{0,1,\dots,  \lfloor\frac{t_{\text{max}}}{\delta t}\rfloor \}$. We represented the sequences of ICIs by sequences of dICIs from that finite set. Note that any ICI value above $t_{\text{max}}$ is mapped to the dICI $\lfloor\frac{t_{\text{max}}}{\delta t}\rfloor$ and therefore represents the end of a coda. We set $t_{\text{max}}=1$ (longer than any ICI) and $\delta t=0.05$ throughout the analysis (see \textit{Supplementary Materials} section $3.3.2$ for justification of this choice and section $3.4.3$ for an analysis on the influence of this parameter).

\subsection*{Variable length Markov chains}
We then modeled these dICI sequences using variable length Markov chains (VLMCs). VLMCs provide the large memory advantage of higher-order Markov chains when needed, without the drawback of having too many unnecessary parameters in the model. 

Fitting a VLMC is the process of deciding how much memory is necessary to model specific sequences. The criterion for making this decision is the following: longer sequences are discarded if their distribution of transition probabilities is similar to that of shorter subsequences. This process is often called \textit{context tree estimation} and consists of two steps. 

The first step is to consider $\mathcal{W}_D$ the set of all sequences of maximum length $D$ (which we set to 10) and to assign the following probability distribution $q_w$ to each sequence:
\begin{equation}
    q_w = P(X|w) ,
\end{equation}
that is, the probability of observing a state $x\in \mathcal{X}$ given the sequence $w$.

The second step is to prune the sequences that do not add information. Take two sequences $u,w \in \mathcal{W}_D$, one being the suffix of the other $w$ = $\sigma u$. The information gained $H_w$ by considering the longer sequence can be measured with a weighted Kullback-Leibler (KL) divergence $D_{KL}$ \cite{kullback1951information}. The longer memory sequence $w$ is kept only if the information gain is greater than some threshold $K$ \cite{machler2004variable,galves2012context}
\begin{equation}
 \Delta H_w =N(w)D_{KL}(q_w || q_u ) > K 
    \label{eq:info_gain}
\end{equation}
where $N(w)$ denotes the length of sequence $w$. Sequences that satisfy this condition are called \textit{contexts} and sequences that do not are discarded. A VLMC can be defined as the set of these contexts $w$ and their associated probability distribution $q_w$ (see \textit{Supplementary Materials} section $3.1$ for details).

A VLMC can be visualized as a tree by representing each context $w$ by a node and setting the root node as the context of length zero. Contexts that are subsequences of each other are then part of the same branches, which end with the longest contexts. 

\subsection*{Quantitative Comparison of VLMCs}
If two VLMC models $T_1$ and $T_2$ are built over the same finite set of dICIs $\mathcal{X}$, there exists a map $\phi_1: \mathcal{W}_{D} \rightarrow T_1$ that maps any sequence of elements of $\mathcal{X}$ into the longest sequence present in $T_1$, and similarly for $T_2$. This map also induces a map between the probability distributions of $T_1$ and $T_2$. Given two distributions over the same set $\mathcal{X}$, we can measure how different they are with the $KL$ divergence. Therefore, it is possible to define a dissimilarity between $T_1$ and $T_2$ by considering the average $KL$ divergence over all sequences of $T_1$ and their map $\phi_1(T_2) \subseteq T_1$

\begin{equation}
    \begin{aligned}
    d_{KL}(T_1,T_2) &= \frac{1}{|T_1|}\sum_{w\in T_1}D_{KL}\left(q_w || p_{\phi_1(w)}\right) \\
    \end{aligned}
\end{equation} 
Refer to the \textit{Supplmentary Materials} section $3.4$ for a more detailed explanation.

This results in a dissimilarity measure that captures not just the difference in emission distribution but also the structural differences of the associated context trees. 
When comparing the distribution of distances in \cref{fig-validation-dominica}A and \cref{fig-results-pacific}A we performed a \textit{Kolmogorov–Smirnov} test to test if the distances between social units/coda samples of the same clan and distances between social units/coda samples of different clans had come from the same distribution. For every pair, we can reject the hypothesis of the distances coming from the same distribution with $95\%$ confidence. 

\subsection*{Hierachical Clustering of VLMCs}
The dendrograms in \cref{fig-validation-dominica}B and \cref{fig-results-pacific}B were obtained by hierarchical clustering using average linkage on the set of subcoda trees (VLMCs). Since the distance is not symmetric, for agglomerative clustering we considered the symmetric distance:
\begin{equation}
    d_{KL}(T_1,T_2)= \max \left\{d_{KL}(T_1,T_2),d_{KL}(T_2,T_1)\right\} .
\end{equation}

\subsection*{Measuring clan overlap}
\label{sec:clan_overlap}
We used the clan spatial overlap values from \cite{hersh2022evidence}. Briefly, given two clans A and B, and the coda samples associated to them, the amount of geographical overlap of A in B was measured as the fraction of coda samples belonging to clan A that were recorded within 1000 kilometers of at least one repertoire of clan B.
One thousand kilometers is the approximate annual home range span of sperm whales in the eastern tropical Pacific \cite{whitehead2001analysis,whitehead2008movements}.

\subsection*{Statistical Testing}
On \cref{fig-validation-dominica} and \cref{fig-results-pacific} we compare the distributions of distances between subcoda trees of coda samples/social units of the same clan (\textit{within}) and of different clans (\textit{between}). The purpose is to assess whether these distributions originate from the same underlying population. We employ both the \textit{Kolmogorov–Smirnov} test and the \textit{T-}test. The observed \textit{p}-values were well below $0.01$ for all clans. This allows us to confidently reject the hypothesis that there is no difference between the vocal style between different clans. For more information check the \textit{Supplementary Materials} in section $3.4.2$. 

To assess the existence of a relationship between clan overlap and vocal style similarity, we applied an ordinary least squares linear regression model (OLS). 
We show the resulting $p$ values of the OLS statistical test at the bottom left of each plot of \cref{fig:similarity-clan-overlap} along with the observed $r^2$ value.
To assess whether there is true difference between the two cases, we also bootstrapped the linear regression calculation to obtain 95\% confidence intervals for the slopes of the fits, resulting in both negative and positive values in the ID case, but only negative slope values for the non-ID case, thus confirming our interpretation.

\section*{Acknowledgments}
This study was funded by Project CETI via grants from Dalio Philanthropies and Ocean X; Sea Grape Foundation; Rosamund Zander/Hansjorg Wyss, Chris Anderson/Jacqueline Novogratz through The Audacious Project: a collaborative funding initiative housed at TED. TAH was supported by Max Planck Group Leader funding to Andrea Ravignani of the Max Planck Institute for Psycholinguistics. SP was supported by the Polish National Science Center grant UMO-2016/20/W/NZ4/00354 funding to Wlodzislaw Duch of the Nicolaus Copernicus University in Torun.
The Dominica coda dataset originates from The Dominica Sperm Whale Project which was supported by a FNU fellowship for the Danish Council for Independent Research supplemented by a Sapere Aude Research Talent Award, a Carlsberg Foundation expedition grant, a grant from Focused on Nature, two Explorer Grants from the National Geographic Society (all to SG), and supplementary grants from the Arizona Center for Nature Conservation, Quarters For Conservation, the Dansk Akustisks Selskab, Oticon Foundation, and the Dansk Tennis Fond. Further funding was provided by Discovery and Equipment grants from the Natural Sciences and Engineering Research Council of Canada to Hal Whitehead (Dalhousie University) and a FNU large frame grant and a Villum Foundation Grant to Peter Madsen (Aarhus University). The publicly accessible Pacific Ocean sperm whale coda dataset we used in this study emanates from the Global Coda Dialect Project, a consortium of scientists conducting sperm whale acoustics research worldwide. Members of the consortium who contributed to the Pacific Ocean dataset include: Luke Rendell, Mauricío Cantor, Lindy Weilgart, Masao Amano, Steve M. Dawson, Elisabeth Slooten, Christopher M. Johnson, Iain Kerr, Roger Payne, Andy Rogan, Ricardo Antunes, Olive Andrews, Elizabeth L. Ferguson, Cory Ann Hom-Weaver, Thomas F. Norris, Yvonne M. Barkley, Karlina P. Merkens, Erin M. Oleson, Thomas Doniol-Valcroze, James F. Pilkington, Jonathan Gordon, Manuel Fernandes, Marta Guerra, Leigh Hickmott and Hal Whitehead. We are grateful to Scott Baker and Alana Alexander for answering questions about sperm whale genetics.

\bibliographystyle{naturemag}
\bibliography{bib}

\end{document}


\maketitle
\section{Data and Preprocessing} \label{sec:preprocessing}

Sperm whales communicate vocally via \textit{clicks}: short bursts of sound emitted in sequence. These clicks are combined into recognizable patterns called \textit{codas}. Clicking sperm whales were recorded with a hydrophone, and clicks were detected in the resulting audio files by human experts. The data that we used consists of time sequences of inter-click intervals (ICIs), i.e. the times between two consecutive detected clicks---this is equivalent to having a time series of click onsets. 

We used two datasets: the Dominica and Pacific datasets. The Dominica dataset contains 8719 annotated codas recorded in the Atlantic Ocean near Dominica. The codas come from 12 social units grouped into two vocal clans (EC1 and EC2). The Pacific dataset consists of around 23555 codas recorded between 1978 and 2017 in 23 Pacific Ocean locations. The codas were divided into coda samples according to their recording day and each coda sample was assigned a single vocal clan inferred in \cite{hersh2022evidence}. When considering a clan-level analysis all coda samples where used to compute the VLMC models. However when analysing at a coda sample level, we discarded coda samples with less than 200 codas with statistical inference in mind, resulting in a final count of 57 coda samples (17046 codas) for the Pacific.

\begin{figure*}[!h]
    \centering
    \includegraphics[height=6cm]{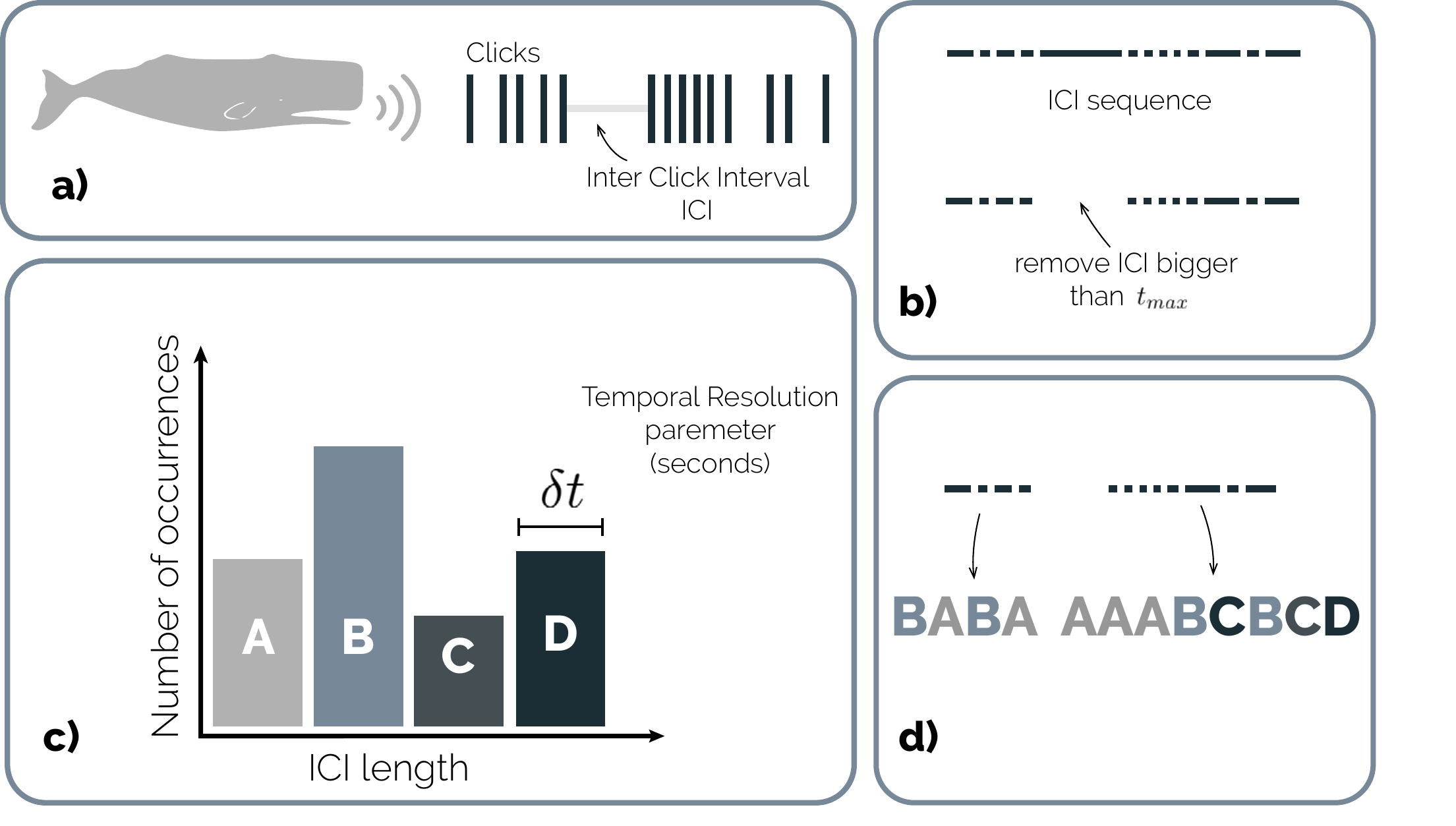}
    \caption{\textbf{Representing sperm whale communication as a sequence of symbols.} \textbf{a} The temporal patterns of click onsets are equivalently encoded as sequences of inter-click intervals (ICIs). \textbf{b} We discard ICI values larger than a threshold $t_{\text{max}}$ seconds, and \textbf{(c)} discretize the others into a finite set of bins of width $\delta t$ seconds. Each bin is then assigned a symbol (here a letter of the Latin alphabet), so that each symbol represents an ICI in a given range. \textbf{d} We then represent sperm whale communication as a sequence of these symbols.}
    \label{fig:preprocessing}
\end{figure*}

To model the ICI sequences in the datasets, we represented the ICIs by a finite set of symbols (or states), in three main steps (\cref{fig:preprocessing}). First, we denoted $X$ a continuous random variable that represents an ICI:
\begin{equation}
X = \text{ICI} .
\end{equation}
Second, we imposed an upper bound $t_{\text{max}}$ on the values taken by $X$. This was to make sure that we are modeling with a finite number of states. Specially, in situations here the maximum number of states have to be known. We set this value to 1 second to be sure that it is longer than any ICI.
\begin{equation}
X_{t_{\text{max}}} = \min\left(\text{ICI}, t_{\text{max}} \right) .
\end{equation} 

Thirdly and finally, we discretized the values of the ICIs into a set of bins, akin to a histogram. We denote $\delta t$ the width of these bins in seconds and call it the \textit{temporal resolution} of the representation. Formally, we define
\begin{equation}
X_{\delta t,t_{\text{max}}} = \left\lfloor\frac{\min\left(\text{ICI}, t_{\text{max}} \right)}{\delta t}\right\rfloor .
\end{equation}
By construction, this defines a discrete random variable that takes values in the finite set $\mathcal{X}=\{0,1,\dots,  \lfloor\frac{t_{\text{max}}}{\delta t}\rfloor \}$. Note that each element of this set represents a range of ICI values of length $\delta t$ seconds. Any ICI value above $t_{\text{max}}$ is mapped to the symbol $\lfloor\frac{t_{\text{max}}}{\delta t}\rfloor$. It thus represents the end of a coda. We set the upper bound to $t_{\text{max}} = 1$ seconds and $\delta t =0.05$ (see \cref{sec:parameters} for details about the choice of values).

We then modeled sperm whale communication sequences $(X_i)_{i\in\mathbb{N}}$. In addition, for clarity, we will denote the elements of $\mathcal{X}$ by letters of the Latin alphabet $A$, $B$, and so on. In terms of terminology, we will also refer to $\mathcal{X}$ as an alphabet, and to its elements as symbols or states, interchangeably.

\subsection{ID and non-ID codas}
Some coda types are considered ID for some clans but non-ID for others. 
The Pacific dataset \cite{hersh2022evidence} has annotations specifying this label (ID or non-ID) for each coda and each clan. 
For each clan, we counted the total number of codas that are ID and the total number of codas that are not (see \cref{tab:count_codas}). 

\begin{table}[htb]
\centering
\caption{Count of Codas}
\label{tab:clanstats}
\begin{tabular}{lrrrr}
\toprule
\textbf{Clan Name} & \textbf{Total} & \textbf{ID} & \textbf{Non-ID} & \textbf{Percentage of ID}\\
\midrule
Four Plus (FP) & 2590 & 587 & 2003 & 22.66\% \\
Palindrome (PALI) & 2124 & 1249 & 875 & 58.8\% \\
Plus One (PO) & 2083 & 1252 & 831 & 60.11\% \\
Regular (REG) & 8289 & 3545 & 4744 & 42.77\% \\
Rapid Increasing (RI) & 1776 & 425 & 1351 & 23.93\% \\
Short (SH) & 5428 & 562 & 4866 & 10.35\% \\
Slow Increasing (SI) & 1265 & 541 & 724 & 42.77\% \\
\midrule
Total & 23555  & 8161 & 15394& 34.65\% \\
\bottomrule
\end{tabular}
\label{tab:count_codas}
\end{table}

We also provide similar statistics for the count of different coda types (see \cref{tab:count_coda_types}).
\begin{table}[htb]
\centering
\caption{Count of Coda Types}
\begin{tabular}{lrrrr}
\toprule
\textbf{Clan Name} & \textbf{Total} & \textbf{ID} & \textbf{Non-ID} & \textbf{Percentage of ID}\\
\midrule
Four Plus (FP) & 72 & 2 & 70 & 2.78\% \\
Palindrome (PALI) & 64 & 9 & 55 & 14.06\% \\
Plus One (PO) & 54 & 6 & 48 & 11.11\% \\
Regular (REG) & 74 & 9 & 65 & 12.16\% \\
Rapid Increasing (RI) & 79 & 2 & 77 & 2.53\% \\
Short (SH) & 85 & 1 & 84 & 1.18\% \\
Slow Increasing (SI) & 59 & 3 & 56 & 5.08\% \\
\midrule
Total & 487 & 32 & 455 & 6.57\% \\
\bottomrule
\end{tabular}
\label{tab:count_coda_types}
\end{table}

\section{The role of memory}
\label{sec:memory}
A coda is a series of clicks emitted in fast succession. 
They have traditionally been identified by practitioners as building blocks for sperm whale communication. 
Whether or not codas themselves are composed by series of smaller collections of clicks is however an open question. 
We asked: are the codas the smallest such blocks, or is there structure at a scale shorter than codas---but longer than the individual clicks that constitute them? 
In order to answer this, we modeled sperm whale communication as higher-order Markov chains, that is, Markov chains with a memory $h$ larger than or equal to one (but of fixed length)---see \cref{sec:memory} for details.  
In other words, we assume that the probability of observing an ICI within given range---here referred to as symbol or state---depends on the $h$ previous states. 
For a given $h$, we fit this Markov model to the data by estimating the transition probabilities from sequence of $h$ state to any other state. 

\begin{figure*}[htbp]
  \centering
    %
    \includegraphics[width=0.9\linewidth]{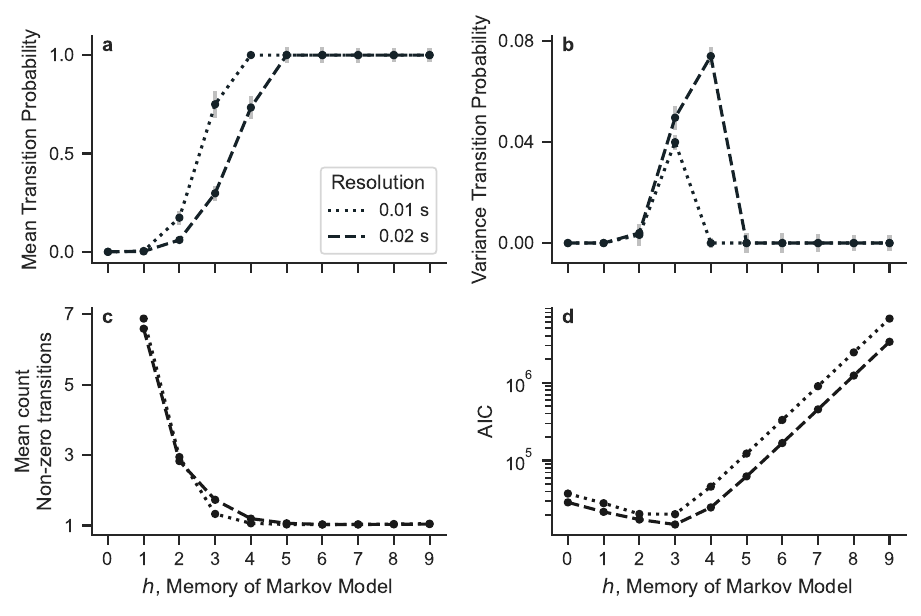}
  \caption{  \textbf{Fixed length Markov model bifurcates as a function of memory length $h$.} We show (\textbf{a}) the mean, and (\textbf{b}) the variance of the transition probabilities as a function of $h$. We also show (\textbf{c}) the number of non-zero transition probabilities per state, and (\textbf{d}) the AIC. The dashed and dotted lines represent two different temporal resolutions (i.e., bin size used during Preprocessing). There is a bifurcation around $h \approx 3$ that suggests sub-coda structure.}
  \label{fig:markov_results}
\end{figure*}

The results are summarized in \cref{fig:markov_results}, for a range of memory values $h$, and for two temporal resolutions for the binning of the ICIs.
We note that there is a bifurcation between two different behaviors around $h \approx 3$: transition probabilities go from very low (approximately random transitions) to very large (almost deterministic transitions). 
Indeed, for $h<3$, these probabilities are very low [\cref{fig:markov_results}(a)] and all similar [\cref{fig:markov_results}(b)]. All possible next states are equally likely, but not very likely: this indicates underfitting. 
On the contrary, for $h>3$, the transition probabilities are all close to one and all similar. 
Moreover, only a few of them are non-zero [\cref{fig:markov_results}(c)]. 
Given a sequence longer than three, only one next state can be observed: this indicates overfitting. 
Finally, for $h \approx 3$, transition probabilities are heterogeneous: their average is between 0 and 1, their variance exhibits a peak, and more than one state can potentially be observed next. Moreover, the Akaike Information Criterion (AIC) displays a minimum around that value of the memory, which indicates that it provides a good trade-off between variance of the data explained and the number of parameters needed for the model. 

This transition around $h \approx 3$ suggests that there is structure at that level of memory, which is shorter than most coda types. 
This motivates our search for structure within codas. 
However, fixed-memory Markov Chains do not allow for different configurations to have different levels of memory, which leads to variable length Markov Chains (described in the next section). 

\section{Variable Length Markov Chains} \label{sec:vlmc}

Some states can be predicted with more or less memory of past states than others. This observation is the base motivation for introducing variable length Markov Chains (VLMCs) which go beyond the fixed-memory limit of traditional Markov chains. 
Take for example a state $X_2$ that has the same probability of occurring knowing the last two states ($x_0x_1 \in \mathcal{X}^2$) or only the last state ($x_1 \in \mathcal{X}^1$):
\begin{equation}
    P(X_2 | x_0,x_1 ) = P(X_2|x_1) .
\end{equation} 
In this case, a shorter memory ($h=1$) is sufficient and we do not need a longer one ($h_2$).

In practice, VLMCs bypass the necessity of having $(n-1)n^h$ parameters by allowing states to have unequal lengths (memory). 
Smaller lengths are preferred whenever the additional memory does not significantly change the distribution of transitions to the next possible states.

\begin{figure*}[htb]
    \centering
    \includegraphics[height=5cm]{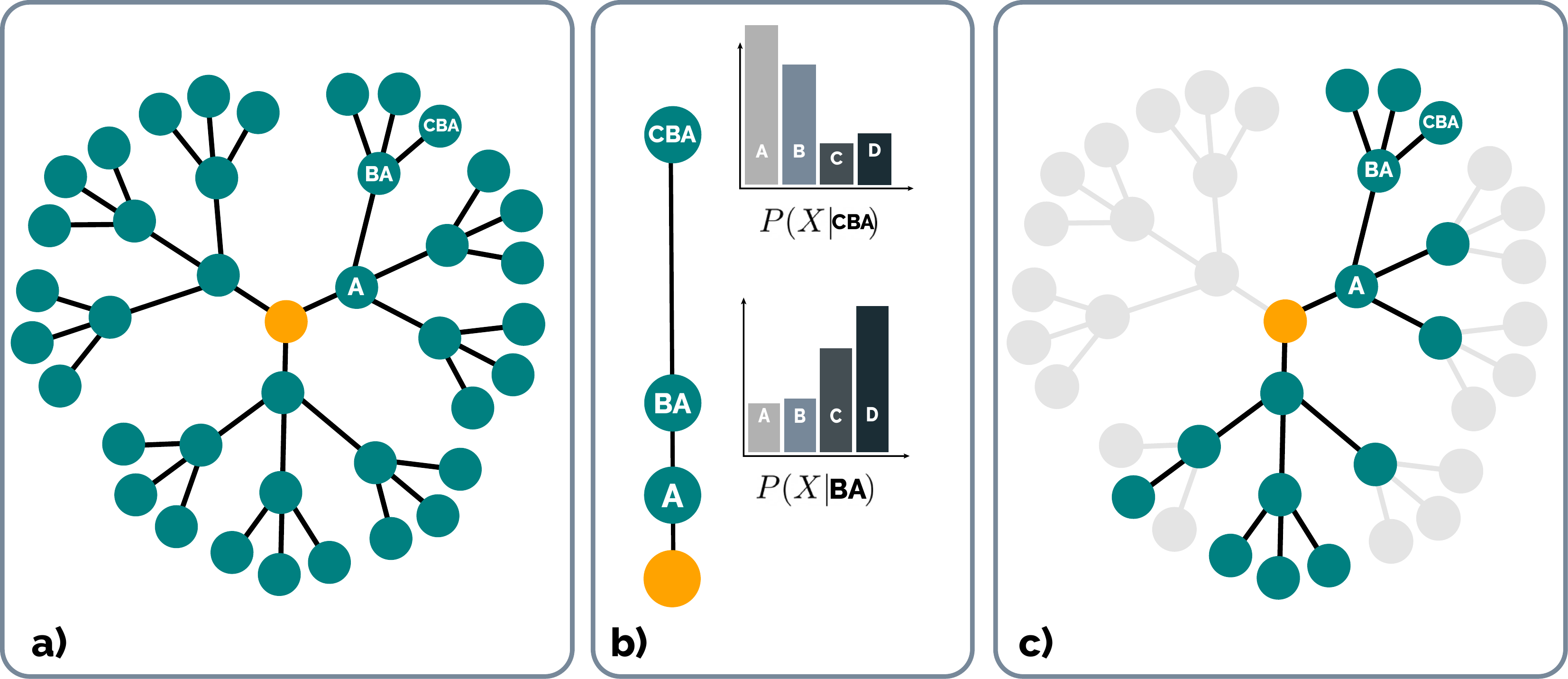}
    \caption{  Steps for fitting a VLMC from a sequence. Step 1 (a): Construction of the full (\textit{or saturated}) suffix tree up to maximum depth $D$. Step 2 (b): Assigning a probability measure to every element of the tree. Step 3 (c) Pruning of nodes that carry the same information content as parents according to \cref{eq:threshold}. In this example, context $CBA$ is relevant because it changes the distribution of transition probabilities with respect to its parent $BA$ and therefore it is not pruned. {\rev The orange node represents the root node.}}
    \label{fig:vlmc_example_2}
\end{figure*}

\subsection{Building a VLMC}
For a Markov model of fixed order $h$, the set of possible states $\mathcal{X}^h$ is composed of all possible sequences of length $h$. For a VLMC, however, states can be sequences of arbitrary length. 
The set of possible states is thus a subset of the set of all sequences that can be built from the alphabet $\mathcal{X}$, including the empty sequence $\mathcal{X}^0 =\varnothing$. Let $\mathcal{W}$ denote this set.
\begin{equation}
    \mathcal{W} = \bigcup_i^{\infty} \mathcal{X}^i
\end{equation}
In this project, because codas are typically constructed from a small number of clicks, we only consider \emph{finite} length sequences (see \cite{peffy2010variable} for non-finite VLMCs). 
In practice, we choose a \emph{maximum} memory allowed $D$, which we set to $D=10$, much larger than the typical coda length.    
\begin{equation}
    \mathcal{W}_D = \bigcup_i^{D=10} \mathcal{X}^i
\end{equation}


Fitting a VLMC is the process of finding some subset $\mathcal{L} \subseteq \mathcal{W}_D$ where the elements satisfy the condition: shorter states are preferred if their distribution of transition probabilities is similar to their longer length equivalents. This is generally called context tree estimation in the literature \cite{gustafsson2021fast}.

\textbf{Probabilising the tree}
We start with $\mathcal{W}_D$ for some $D$ which we take to be equal to $10$. To each element of $w\in \mathcal{W}_D$ we assign a probability distribution $q_w$ over the set $\mathcal{X}$ as the probability of observing a state $x\in \mathcal{X}$ given a previous sequence $w$. 
\begin{equation}
    q_w = P(X|w)
\end{equation} Where $P$ denotes the likelihood estimation computed as
\begin{equation}
    q_w(X=x) = P(X=x|w) = \frac{N(wx)}{\sum_{c\in\mathcal{X}}N(wc)}
    \label{eq:likelihood}
\end{equation} where $N(w)$ the number of occurrences of the sequence $w$.

\textbf{Pruning the tree}
Given two sequences $u,w \in \mathcal{W}_D$, we say that $u$ is a suffix of $w$ if $w$ = $\sigma u$ for some other sequence $\sigma$ of length $\geq 1$. If $\sigma \in \mathcal{X}$ we say that $u$ is a parent of $w$. 
That is, $u$ is a parent of $w$ if $u$ is a suffix of $w$ and $w$ is longer by only one letter.

In an intuitive way, $u$ is a parent of $w$ if $w$ ``looks into the past'' one step further than $u$. At the core of the VLMC is measuring the information gain in using the longer memory $u$ instead of its shorter memory parent $w$. If this information gain is not sufficient, then we discard the longer memory $u$. We measure the information gain in using the longer memory $w$ instead of $u$ with a weighted Kullback-Leibler (KL) divergence $D_{KL}$ \cite{kullback1951information}:

\begin{equation}
    \begin{aligned}
 \Delta H_w &=N(w)D_{KL}(q_w || q_u )\\
 &= N(w)\sum_{x\in \mathcal{X}} P(x | w) \log{\frac{P(x|w)}{P(x|u)}}
    \end{aligned}
\end{equation}

The longer memory sequence $w$ is kept if and only if the information gain is greater than some threshold $K$ \cite{machler2004variable,galves2012context}. 
Refer to \cref{sec:parameters} for a discussion on the value of $K$. 
The set of all sequences that respect the above threshold are called \textit{contexts}, and denoted by $T$:
\begin{equation}
    T = \{ w \in \mathcal{W}_D |\Delta H_w > K \}
    \label{eq:threshold}
\end{equation}

A VLMC is the the Markov model with the set of states:
\begin{equation}
    \mathcal{L} = \{wx \hspace{0.5em}| \hspace{0.5em} t \in T, x\in \mathcal{X}\}
\end{equation} and with transition probabilities defined by $q_w$ for $w \in T$.

\subsection{Model Selection}
When modeling a process $(X_i)_{i\in \mathbb{N}}$ with a Markov model, the memory length $h$ controls the trade-off between complexity and error. 
Higher memory values tend to result in models that generalize poorly. On the other hand, lower values of $h$ fail to capture the patterns, resulting in a uniformly random model.

To choose an appropriate value of $h$, it is common to employ some statistic that measures the trade-off between precision in prediction and the number of parameters. A model with high predictability and a low number of parameters is favored. There is a wide range of metrics \cite{chang2019predictive} and one of the most widely used is the AIC \cite{akaike1974new,tong1975determination}:
\begin{equation}
    \text{AIC} = 2k - 2\log{\hat{P}\bigl((X_i)_{i \in \mathbb{N}}|\mathcal{M}_h\bigr)} .
\end{equation}
$\hat{P}\bigl((X_i)_{i \in \mathbb{N}}|\mathcal{M}_h\bigr)$ is the maximum likelihood of the sequence $(X_i)_{i \in \mathbb{N}}$ given the Markov model $\mathcal{M}_h$ with memory length $h$ and $k$ parameters (transition probabilities). The best model is indicated by the lowest AIC.

\subsection{Parameter sensitivity}
\label{sec:parameters} 
\subsubsection{Information gain threshold $K$ } 
\begin{figure*}[!h]
    \centering
    \includegraphics[height=7cm]{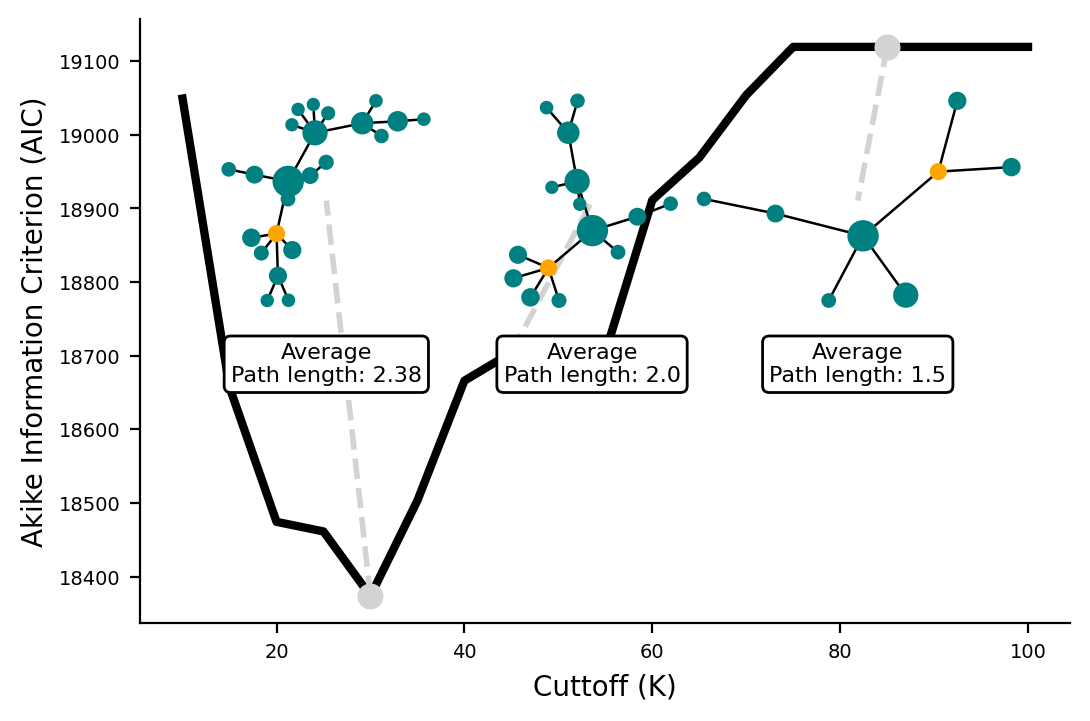}
    \caption{\textbf{Selecting the optimal information cutoff $K$.} The AIC for different VLMC models with respect to the cutoff value $(K)$ used to build them. 
    Above are some context trees associated with three different VLMCs. 
    The trees decrease in size for the threshold increases since the information gain has to be greater for a context to be ``accepted'' in the tree. 
    This effect is also visible in the average path length which can be seen as the average memory of the model.}
    \label{fig:aic_cutoff}
\end{figure*}

The threshold value $K$ represents the minimum information gain necessary to increment the memory of a given context by one. This value ultimately influences the depth and shape of the VLMC model. Low thresholds result in deep trees with many parameters and are prone to overfitting, whereas small threshold values cause trees to not expand past low values of memory and potentially fail to capture statistical dependencies (\cref{fig:aic_cutoff}).

The best value of $K$ should be the one that outputs the optimal VLMC model (i.e., the one that minimizes the AIC).
\begin{equation}
    \arg\min_{T\subseteq\mathcal{W}_D}\text{AIC}(T)
\end{equation} However, this search is done over the entire set of possible context trees. The problem of estimating the optimal context tree is an ongoing research area although many good methods have been proposed \cite{galves2012context}.

For some threshold $K$. The above dissimilarity is an expression of differences of deviances and as such follows an asymptotic $\frac{1}{2}\chi^2$ distribution \cite{machler2004variable} with $|\mathcal{X}|-1$ degrees of freedom. As such we can set the $K$ thresholds to represent quantiles of a $\chi^2$ distribution. We use the 0.95 quantile, meaning we keep the child whose additional memory exceeds the value:

 \begin{equation}
     K = \frac{1}{2}\chi^2_{|\mathcal{X}|-1, 0.95}
 \end{equation}

\subsubsection{Temporal resolution $\delta t$} 
The temporal resolution $\delta t$ denotes the scale at which we discretize the continuous, absolute ICI values into bins. 
A small value might provide differentiation between clicks, but also burdens the VLMC models by increasing the number of parameters and states.

To select the most appropriate resolution parameter, one might be tempted to compare the AIC obtained from different VLMC models extracted from data at different resolutions. However, that is not possible since models fit on different data are not really comparable. 
Imagine the extreme case with a time resolution so large that all clicks are mapped to the same discrete symbol: any model fit on this data would achieve an optimal AIC value.

\begin{figure*}[!h]
    \centering
    \includegraphics[height=6cm]{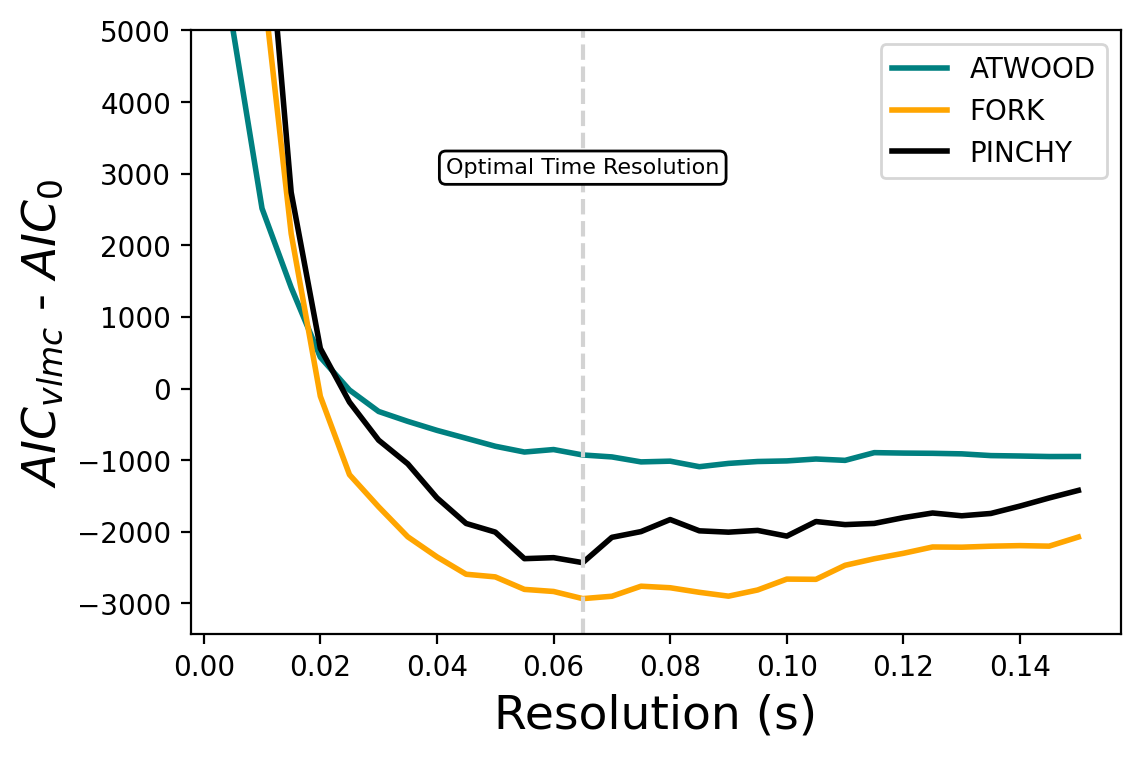}
    \caption{\textbf{Selecting the optimal temporal resolution.} The temporal resolution parameter $\delta t$ for the preprocessing steps is chosen by subtracting the AIC of the 0th order memory Markov model from the AIC of a VLMC model, which encodes how well the model predict versus how easy it is to predict the data. We use this setup because the AIC alone is not sufficient, because  the actual fitted dataset changes with the resolution parameter, and thus models with different $\delta t$ are not directly comparable. The minimum value is the best resolution parameter although any value between $0.01$ and $0.1$ seems acceptable. We show this for three individual sperm whales (colors) from the Dominica dataset.}
    \label{fig:temporal_resolution}
\end{figure*}

In our case, we compare the AIC obtained from the VLMC model with the AIC obtained from a fixed length Markov model of order $0$ fitted to the same aggregated data. 
The intuition behind this approach is that the zero length Markov model represents how ``easy'' it is to predict the data. 
The best resolution would be the one where the difference between the AIC of our fit VLMC is the biggest when compared to the $0^{th}$ order Markov model 
(\cref{fig:temporal_resolution}).

\subsection{Quantitative comparison of VLMCs}
\label{sec:distance}
The KL divergence is one of the most used methods for measuring statistical dissimilarity between two distributions, mostly due to its connections to information-theory. 
Being a generative model, a VLMC is not a single probability distribution, but a set of distributions $q_w$ one for each context $w \in T$. 

Given two VLMC models $T_1$ and $T_2$ over the same alphabet $\mathcal{X}$. Let $p$ and $q$ be their associated sets of transitions distributions, respectively. We define the divergence $d_{KL}(T_1,T_2)$ between them as the average KL divergence between the set of associated transition distributions.

However, there may not be a one-to-one map between $q_w$ and $p_w$. In fact, more often than not $T_1$ and $T_2$ have a different number of contexts. As such, for every $w \in T_1$ we associate $u \in T_2$ where $u$ is the longest suffix of $w$ that belongs to $T_2$. This results in a dissimilarity measure that captures not just the difference in emission distribution but also the structural differences of the associated context trees:

\textbf{Divergence}
Given two VLMCs $(T_1, (q_w)_{w\in T_1})$ and $(T_2,(p_w)_{w\in T_2})$ built over the finite alphabet $\mathcal{X}$ we define the distance:

\begin{equation}
    \begin{aligned}
    d_{KL}(T_1,T_2) &= \frac{1}{|T_1|}\sum_{w\in T_1}D_{KL}\left(q_w || p_{\overleftarrow{\text{pref}}(w)}\right) \\
    &= \frac{1}{|T_1|}\sum_{w\in T_1}\sum_{x\in \mathcal{X}} q_w\log{\frac{q_w}{p_{\overleftarrow{\text{pref}}(w)}}}
    \end{aligned}
\end{equation}
Where $\overleftarrow{\text{pref}}(w)$ denotes the longest suffix of $w \in T_1$ that belongs to $T_2$.
\begin{equation}
    \overleftarrow{\text{pref}}: T_1 \rightarrow T_2
\end{equation}

\begin{figure*}[!h]
    \centering
    \includegraphics[height=3cm]{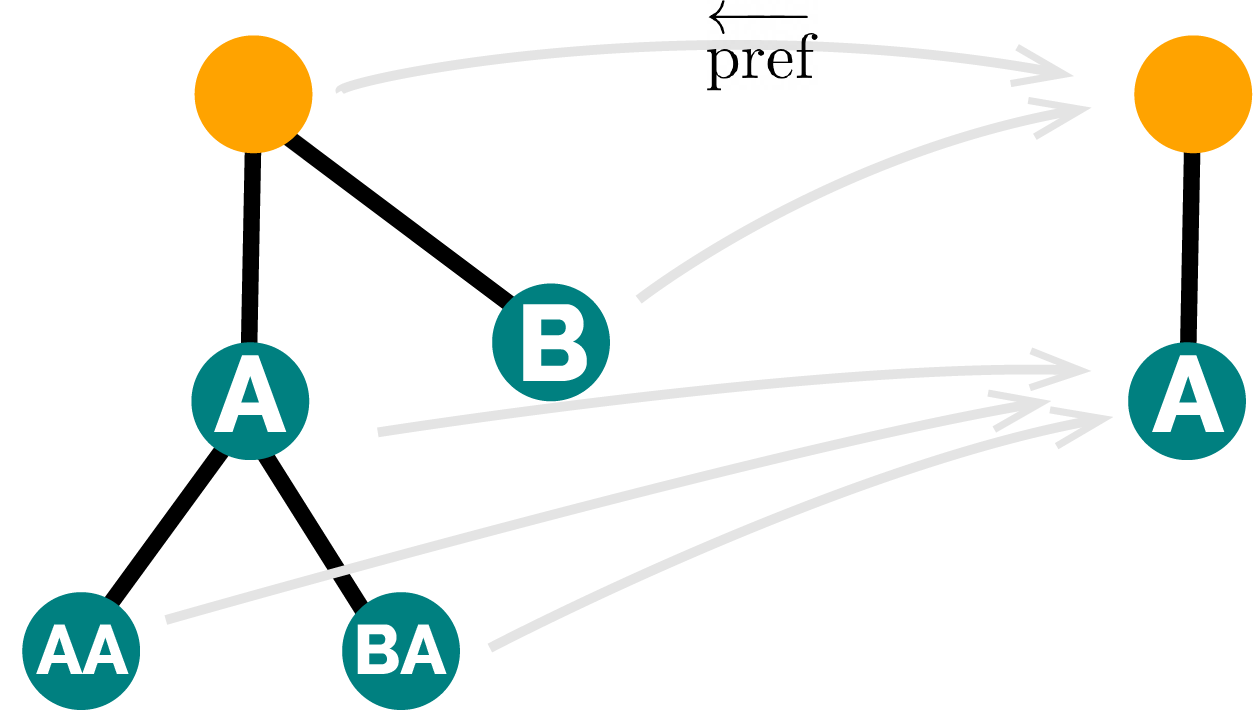}
    \caption{\textbf{Measuring dissimilarity between two VLMC models.} The two trees represent two VLMCs models built over the same alphabet $\mathcal{X}$. We measure their dissimilarity or distance with the KL divergence.}
    \label{fig:coda_representation}
\end{figure*}

\subsubsection{Statistical testing on distribution of the distances}
\begin{figure*}[!h]
    \centering
    \includegraphics[width=\textwidth]{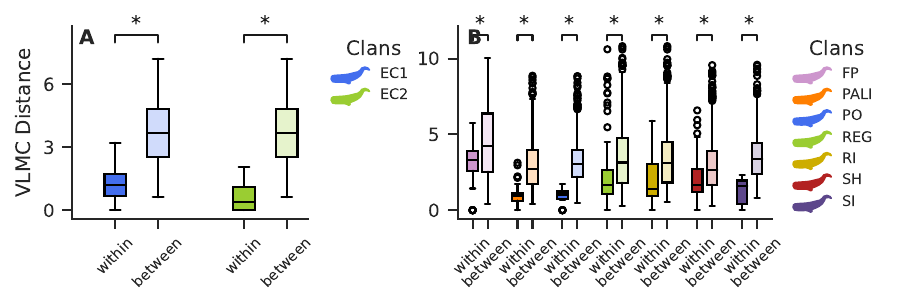}
    \caption{\textbf{Comparison of the distribution of distances between VLMC models, with statistical significance}. Each color represents a clan of a given dataset: \textbf{A)} Pacific, and  \textbf{B)} Dominica. The \textit{within} distribution represents the distance between VLMC models fit on elements of the same clan. The \textit{between} represent the distance between VLMC models of different clans.}
    \label{fig:statistical_distances}
\end{figure*}

We fit a VLMC model on coda samples/social units from different clans on both the Pacific and Dominica dataset.
For each clan we compute the distances between all VLMC models belonging to that clan (\textit{within}) and between the models of the clan to the ones belonging to other clans (\textit{between}) (\cref{fig:statistical_distances}).
On each pair (\textit{within/between}) we tested the distributions to check if they both came from the same populations. We employed both the \textit{Kolmogorov-Smirnov} test and the \textit{T-}Test.
We also measured the effect size using \textit{Cohen's} method. The resulting statistics for both datasets can be found in the tables below.

\begin{table}[htb]
\centering
\caption{\textbf{Distance distribution statistics}. Table with the \textit{p-}values and the effect size corresponding to the comparison of the \textit{within/between} distance distributions of coda samples from the Dominica clans (top) and the Pacific clans (bottom), under the Kolmogorov-Smirnov (K-S test) and the T-Test.}
\begin{tabular}{lccc}
\toprule
\textbf{Clan} & \textbf{K-S test} & \textbf{T-test} & \textbf{Effect Size} \\
\midrule
EC1 & $1.5\times10^{-16}$ & $1\times10^{-24}$ & $-2.4$ \\
EC2 & $0.0019$ & $0.0006$ & $-1.9$ \\
\midrule
Four Plus & $1.7\times10^{-7}$ & $6.2\times10^{-6}$ & $-0.67$ \\
Palindrome & $6.7\times10^{-17}$ & $6.7\times10^{-13}$ & $-1.3$ \\
Plus One & $6.9\times10^{-21}$ & $2.4\times10^{-14}$ & $-1.6$ \\
Regular & $4.2\times10^{-27}$ & $7.2\times10^{-28}$ & $-0.76$ \\
Rapid Increasing & $0.00016$ & $0.0005$ & $-0.72$ \\
Short & $6.5\times10^{-14}$ & $9.9\times10^{-18}$ & $-0.67$ \\
Slow Increasing & $1.6\times10^{-9}$ & $1.6\times10^{-7}$ & $-1.4$ \\
\bottomrule
\end{tabular}
\label{tab:statistical_difference}
\end{table}

\subsubsection{Non-ID results by coda type}
In this section we repeat the approach on comparing the geographical clan overlap with the VLMC distance on non-ID codas.
In contrast to the main text, we segment each set of codas by coda type and note the slope, the p-value of the Pearson correlation and the p-value according to a Spearman correlation, \cref{tab:coda_types}.
Just as the main text, when segmenting by coda type we observe that the vast majority of correlations is negative, i.e., geographically overlapped clans have a more similar communication.
However, although most correlations are negative, only a small portion is significant.
It is important to take into consideration that the amount of data used to fit each VLMC model is considerably reduced given the extra segmentation.
Furthermore, coda types that were uttered exclusively by only two clans were also omitted as it is always possible to draw a line between two points, and thus a linear analysis makes little sense. 
\begin{figure*}[htb]
    \centering
    \includegraphics[width=0.65\linewidth]{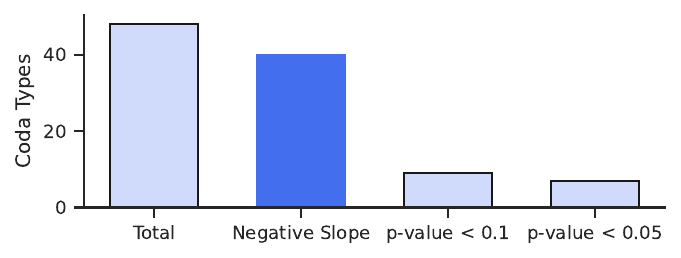}
    \caption{\textbf{VLMC distance and geographical overlap relation by coda type.} each bar represents a number of unique coda types. Note that almost all coda types have a negative correlation with geographical overlap, although only a small portion is significant. The p-values shown are with respect to the Pearson correlation,  \cref{tab:coda_types}}
    \label{fig:coda_types}
\end{figure*}

\begin{table}[htb]
\centering
\caption{\textbf{Geographical overlap and subcoda distance by coda type}. Results of comparing the geographical overlap and VLMC distance segmented by coda types. Negative correlations are highlighted as well as p-values that are below 0.05. ``Number of Clans" represents how many clans were compared.}
\begin{tabular}{ccccc}
\toprule
Coda Type & Slope & Pearson-P & Spearman-P & Number of Clans \\
\midrule
33 & \textbf{-0.353} & 0.864 & 0.787 & 3 \\
34 & \textbf{-0.944} & 0.257 & 0.050 & 4 \\
35 & \textbf{-6.022} & \textbf{0.031} & \textbf{0.032} & 5 \\
42 & \textbf{-4.098} & \textbf{0.004} & \textbf{0.010} & 7 \\
43 & \textbf{-0.615} & 0.596 & 0.573 & 4 \\
45 & \textbf{-1.327} & 0.266 & 0.207 & 5 \\
47 & \textbf{-5.808} & 0.628 & 0.700 & 3 \\
48 & 9.818 & 0.258 & 0.257 & 3 \\
49 & 10.481 & 0.371 & 0.538 & 3 \\
51 & \textbf{-0.939} & 0.634 & 0.695 & 6 \\
52 & \textbf{-1.663} & \textbf{0.002} & \textbf{0.021} & 7 \\
53 & \textbf{-3.341} & \textbf{0.017} & 0.075 & 6 \\
55 & 0.522 & 0.909 & 0.752 & 4 \\
57 & \textbf{-3.480} & 0.410 & 0.511 & 4 \\
58 & \textbf{-0.896} & 0.755 & 0.623 & 3 \\
61 & \textbf{-1.512} & \textbf{0.074} & \textbf{0.006} & 7 \\
64 & \textbf{-1.651} & 0.061 & 0.276 & 5 \\
66 & \textbf{-2.482} & 0.138 & 0.062 & 7 \\
67 & \textbf{-2.782} & 0.172 & 0.269 & 6 \\
71 & \textbf{-1.994} & 0.870 & 0.518 & 3 \\
72 & \textbf{-0.041} & 0.979 & 0.856 & 7 \\
73 & \textbf{-1.334} & 0.337 & 0.329 & 3 \\
76 & \textbf{-0.981} & 0.681 & 0.829 & 7 \\
77 & \textbf{-4.093} & 0.141 & 0.093 & 5 \\
79 & \textbf{-1.116} & 0.268 & 0.142 & 5 \\
82 & \textbf{-0.847} & 0.265 & 0.266 & 3 \\
83 & 0.908 & 0.578 & 0.284 & 6 \\
85 & \textbf{-1.473} & \textbf{0.044} & \textbf{0.032} & 4 \\
86 & \textbf{-0.089} & 0.845 & 0.600 & 3 \\
88 & 2.831 & \textbf{0.000} & \textbf{0.000} & 7 \\
91 & \textbf{-2.662} & 0.254 & 0.387 & 6 \\
93 & 0.613 & 0.597 & 0.618 & 6 \\
310 & \textbf{-4.118} & 0.320 & 0.193 & 4 \\
311 & \textbf{-1.554} & 0.170 & \textbf{0.016} & 6 \\
312 & \textbf{-1.275} & 0.325 & 0.069 & 6 \\
313 & \textbf{-3.087} & 0.475 & 0.186 & 4 \\
315 & \textbf{-1.189} & 0.593 & 0.700 & 3 \\
411 & 0.489 & 0.711 & 0.795 & 4 \\
412 & \textbf{-2.092} & 0.578 & 0.700 & 3 \\
414 & \textbf{-2.269} & \textbf{0.012} & \textbf{0.006} & 7 \\
511 & \textbf{-3.246} & \textbf{0.038} & 0.059 & 7 \\
513 & \textbf{-1.806} & 0.174 & 0.107 & 7 \\
514 & \textbf{-0.449} & 0.869 & 1.000 & 4 \\
610 & \textbf{-1.794} & 0.145 & 0.123 & 7 \\
612 & 5.633 & 0.486 & 0.618 & 3 \\
613 & \textbf{-0.426} & 0.907 & 0.397 & 3 \\
1011 & \textbf{-1.081} & 0.539 & 0.660 & 5 \\
1012 & \textbf{-0.966} & 0.696 & 0.342 & 4 \\
\bottomrule
\label{tab:coda_types}
\end{tabular}
\end{table}

\subsubsection{Stability under different resolutions}
\begin{figure*}[!htb]
    \centering
    \includegraphics[width=0.7\linewidth]{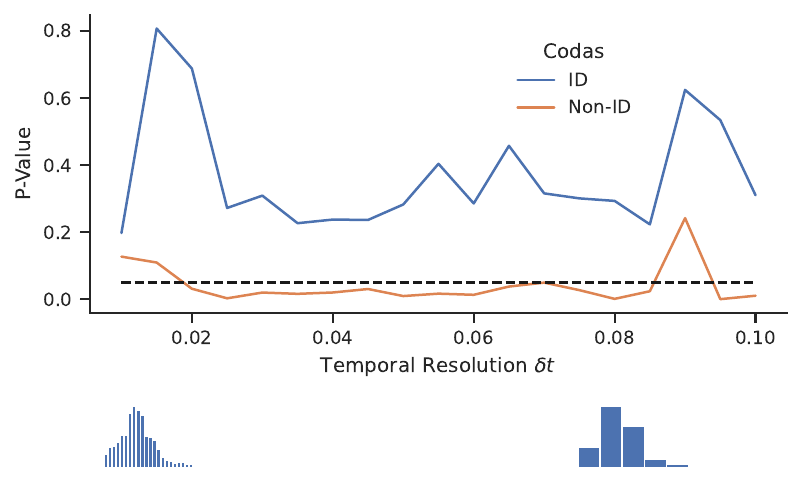}
    \caption{\textbf{Results are significant for different resolutions}. The dotted line represents the $0.05$ confidence value. When varying the time resolution used for fitting the VLMCs we observe that our main results persist: ID codas are not related with geographical overlap while non-ID codas always are. The bottom row shows two illustrations of the bin sizes (and subsequently the number of different states) resulting from the preprocessing using $0.01$ and $0.1$ seconds time resolution respectively.}
    \label{fig:stability}
\end{figure*}

We also show that our results about the effect of sympatry on non-ID coda vocal styles hold for different values of the time resolution.
That is, that the parameter preprocessing steps and the method parameters have little to no effect on the fundamental results of our approach.
In \cref{fig:stability} we repeat the analysis of the main text.
We compare the geographical overlap with the distance between the VLMC of the pacific clans on both non-ID and ID codas.

We observe that regardless of the time resolution used in the method (for discretizing the continuous ICIs into discrete ICIs), our results hold.
That is, there is never a significant correlation between overlap and ID codas and that there is always a negative correlation between clan overlap and non-ID codas.

\subsection{Dependence on Coda Type}

\begin{figure*}[!htb]
    \centering
    \includegraphics[width=0.9\linewidth]{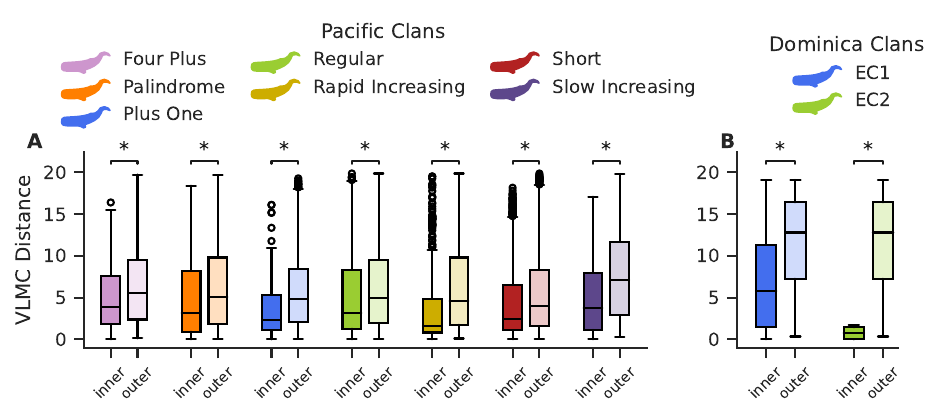}
    \caption{\textbf{Clans vocalize different coda types in a distinctive manner} Boxplots for the distances between VLMC's segmented by coda type and clan (Pacific \textbf{A} and Dominica \textbf{B}). We observe that VLMC fit on different coda types but from the same clan {(\rev inner) are} more similar than when compared with across clans {(\ref outer)}.}
    \label{fig:vlmc_by_type}
\end{figure*}

\begin{figure*}[!htb]
    \centering
    \includegraphics[width=0.9\linewidth]{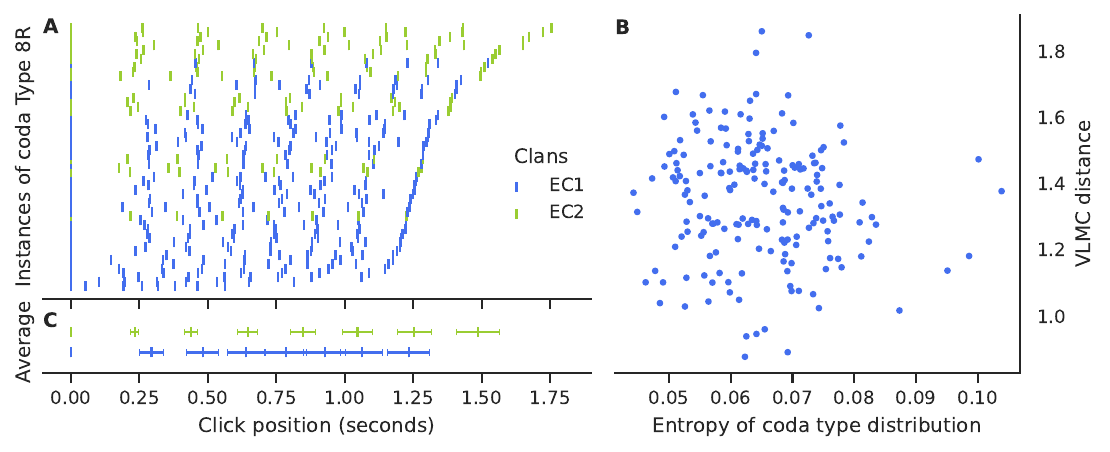}
    \caption{\textbf{A) and C) Rhythmic variations within coda types}. All codas of type \emph{8R} that were vocalized by whales of the Dominican clan EC1 (blue) and EC2 (green), along with the average position and variation over all vocalizations. \textbf{B) No correlation with coda type distribution}. Comparison between VLMC distance and KL-Divergence (entropy) between coda type distributions on the data used to fit the VLMC.}
    \label{fig:type_and_stype}
\end{figure*}

In this section we provide results highlighting the lack of correlation between VLMC (subcoda tree) similarity and coda type distribution. First, rythmic variations on how each coda type is \emph{constructed} are present and are indicative of the clan (Fig. \ref{fig:type_and_stype} and Fig. \ref{fig:vlmc_by_type}). For example, the way the clan EC1 vocalizes codas of type \emph{8R} is significantly different from the clan EC2 (Fig. \ref{fig:type_and_stype}).

In fact one can fit a VLMC on each coda type and compare each VLMC (segmented by coda type and clan) between both elements of the same clan and elements of different clans.
We observe that there is a statistically significant difference between the distances of VLMC from different clans and VLMC of the same clan (Fig. \ref{fig:vlmc_by_type}).
This indicates that whales vocalize different coda types in a clan-distinctive manner.
Which also point to an independence between vocal style and coda type distribution.

\subsubsection{Comparing Dendrograms}
Using the distance between the VLMC trees it is possible to create a hierarchical plot of the coda samples. One can find it beneficial to compare our resulting hierarchical plot with the clan labels from \cite{hersh2022evidence} where the authors group the whales by coda type usage and divide the Pacific clans into the aforementioned 7 clans.
However, comparing a dendrogram with a realized set of labels is not trivial. On the other hand, an effective comparison of two sets of labels can be achieved using the \textit{Adjusted Rand Score}, or other entropy based metrics. 
The Adjusted Rand Score has a value of $0.0$ for random labeling (independent of number of clusters) and $1.0$ for clusters that match perfectly.
The lowest possible score is $-0.5$ for exceptionally disparate clusterings.

\begin{figure*}[htb]
    \centering
    \includegraphics[width=\linewidth]{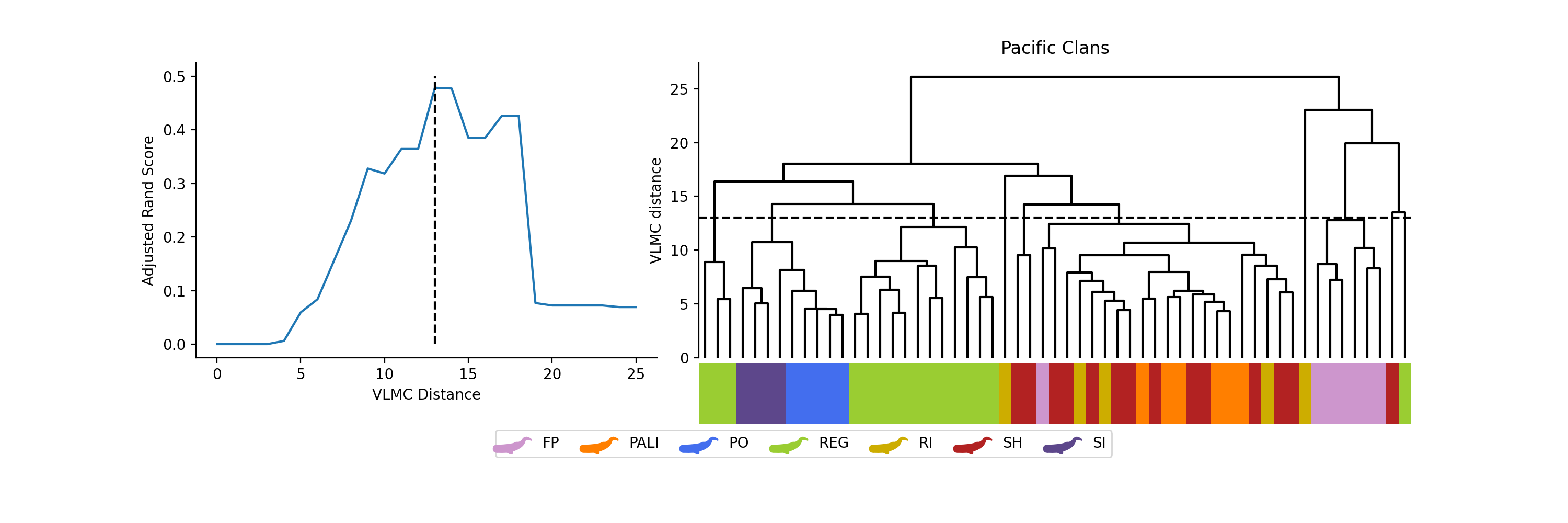}
    \caption{\textbf{Adjusted Rand Score for VLMC clustering dendrogram} Comparison of the dendrogram obtained from the VLMC distance between the trees with the vocal clan labels. The values of the Adjusted Rand Score go from -0.5 to 1.}
    \label{fig:adjusted_rand_score}
\end{figure*}

To obtain two sets of labels we progressively cut the dendrogram obtained by the VLMC and compared the set of labels with the clan labels from \cite{hersh2022evidence}.
At each cut we calculate the adjusted rand score (results in \cref{fig:adjusted_rand_score}). 
We observed a maximum value of $0.5$. This reiterates not only the concordance with the pre-existing vocal usage clans but also emphasizes that vocal style is capturing new information at a different, lower, scale.

\subsection{Confidence Interval on relation between non-id coda style and clan overlap}
\begin{figure*}[!htb]
    \centering
    \includegraphics[width=\linewidth]{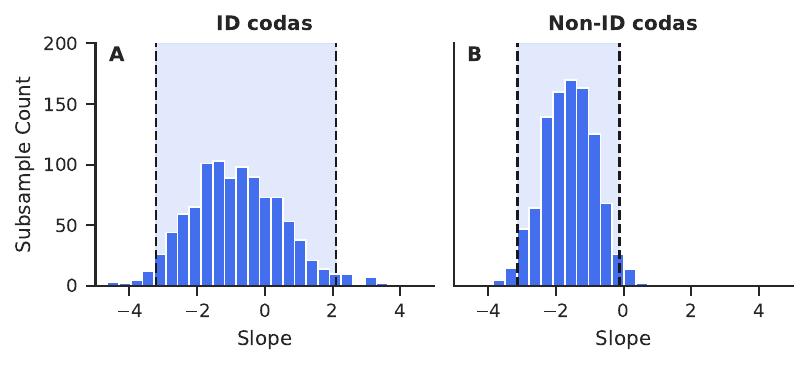}
    \caption{\textbf{Bootstraping regression results of Figure 4.} Resulting distribution of slope values for subsamples of the data for the ID coda case \textbf{A} and the non-ID case \textbf{B}. The shaded area represents the 95\% confidence interval.}
    \label{fig:bootstrap}
\end{figure*}

An interval confidence for the slope for the result in Figure 4 can be achieved by subsampling the data (1000 times) and running the same linear regression analysis on the subsampled data.
From the resulting distribution of regression slopes we observe that the 95\% confidence intervals for the non-ID scenario contains only values with negative slope, while on the ID case, the confidence interval contains both negative and positive values (Fig. \ref{fig:bootstrap}) 

\section{Classification of synthetic codas}
The fixed-length Markov chains described above lack flexibility: a model with large memory $h$ generalizes better but requires estimating an ever-increasing number of parameters. 
For this reason, we then used VLMCs, which combine the best of both worlds by determining the optimal memory needed for each transition individually. 
Essentially, we keep a transition probability with a longer memory $P(X_i | x_j x_k)$ only if it changes the distribution sufficiently compared to a short memory one $P(X_i | x_j)$. 

\begin{figure*}[htb]
    \centering
    \includegraphics[width=0.8\linewidth]{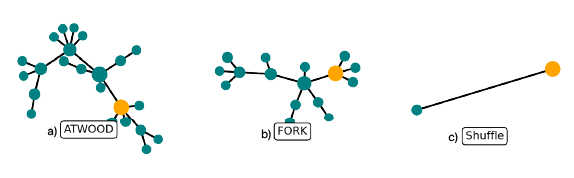}
    \caption{\textbf{VLMCs capture structure of sperm whale communication.} 
    We show context trees associated with the variable length models built for two individual sperm whales: (\textbf{a}) \emph{ATWOOD}, and (\textbf{b}) \emph{FORK}. 
    For comparison, we show (\textbf{c}) the corresponding tree after randomly shuffling ICIs from the same timeseries data used to build (\textbf{a}). Note how the shuffled version does not exhibit any structure. The orange node represents the root node, and the size of each node represents the number of occurrences of the associated context.}
    \label{fig:vlmc_structure}
\end{figure*}

A VLMC can be naturally visualized as a tree, where the concept of order arises from the fact that shorter memory contexts are subsequences of longer ones. In \cref{fig:vlmc_structure}A and B, we show examples of two VLMCs computed from data from a single sperm whale each. The visual structure of these trees can be related to the actual information-theoretic structure of these sperm whales' communication. Indeed, the root node is represented in orange, and nodes that are depth $h$ in the tree (that is $h$ edges away from the root node) represent context (or sequences) of memory $h$. To verify that the structure we observe actually contains information, we need to compare it to the structure of a null model. To do this, we took the same ICI time series used to build the tree from \cref{fig:vlmc_structure}A, and randomly shuffled its ICIs. This way, all temporal information is lost. This results in a tree that has no structure, as shown in \cref{fig:vlmc_structure}C. The VLMC-indicated structure can thus be interpreted as coming from the sperm whale communication. 

\begin{figure*}[htb]
    \centering
    \includegraphics[width=0.75\linewidth]{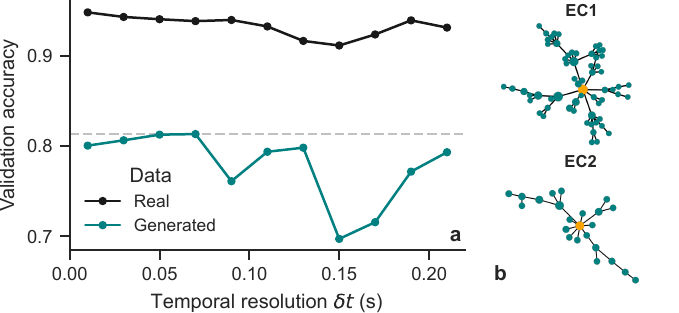}
    \caption{\textbf{Deep learning classifier trained on real communication data generalises well to synthetic data generated by the VLMCs.} We show (\textbf{a}) the accuracy of the classifier on the real data (in black) and the synthetic data (in blue), as a function of the temporal resolution $\delta t$. The dashed grey curve highlights the maximum accuracy of on the generated data. The classifier's task was to identify to which of (\textbf{b}) two clans the data belonged to: EC1 or EC2.}
    \label{fig:vlmc_validation}
\end{figure*}

Having confirmed that our VLMCs capture some communication structure, we ask: What and how much structure does it capture? To answer this, we took advantage of two facts. First, the VLMCs can be used to generate new codas by generating sequences of states that correspond to ICIs.  Indeed, like for any Markov model, we can start from the empty sequence, and start adding suffixes with probabilities defined by the model (see Methods). In other words, we can generate synthetic data. Second, in \cite{bermant2019deep} the authors present an LSTM-based classifier capable of assigning a coda to a specific clan with over $90\%$ accuracy. We trained it on the original ICI data used to build our trees, achieving similar accuracy as shown by the black curve in \cref{fig:vlmc_validation}. To verify how much information our VLMCs capture, we used that trained classifier on the synthetic codas generated with our trees. Remarkably, it classified the generated data with between 70 and 80\% accuracy, depending on the temporal resolution $\delta t$ (blue in \cref{fig:vlmc_validation}). The fairly small difference in accuracy between the real and synthetic data indicates that a large part of the communication structure captured by the classifier in the real data is also captured by our VLMC models. 

\begin{figure*}[htb]
    \centering
    \includegraphics[width=0.75\linewidth]{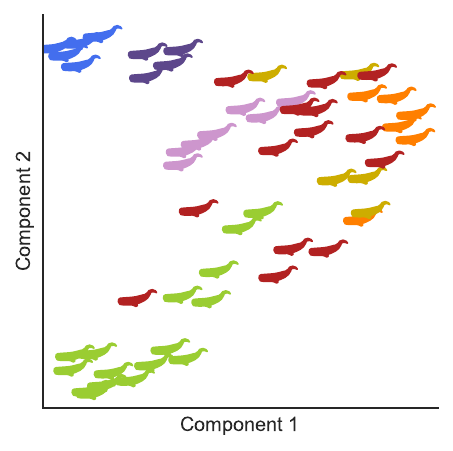}
    \caption{\textbf{Vector Space embedding using UMAP} Embedding of all the Pacific Ocean clans according to their VLMC distance into two dimensions using UMAP.}
    \label{fig:umap}
\end{figure*}

\begin{table}[h!]
\centering
\rev
\caption{ Clan spatial overlap values for Pacific Ocean sperm whale clans as calculated in \cite{hersh2022evidence}. The matrix gives the proportion of the row clan’s repertoires that were recorded within 1,000 km of at least one of the column clan’s repertoires. 1,000 km was chosen because it is the approximate annual home range of eastern tropical Pacific sperm whales (Whitehead, 2001; Whitehead et al., 2008). The matrix is asymmetric because a clan found in only one part of the Pacific Ocean might overlap completely with a clan that spans the Pacific Ocean, while the inverse is not true.
}
\begin{tabular}{|l|l|l|l|l|l|l|l|}
\hline
\textbf{Clan}           & \textbf{FP} & \textbf{RI} & \textbf{P} & \textbf{PO} & \textbf{R} & \textbf{S} & \textbf{SI} \\ \hline
\textbf{Four-Plus (FP)}      & 1                  & 0.500                    & 0.385               & 0.308             & 0.885           & 0.962         & 0.308                   \\ \hline
\textbf{Rapid Increasing (RI)} & 0.368              & 1                        & 0.263               & 0.316             & 0.316           & 0.842         & 0.316                   \\ \hline
\textbf{Palindrome (P)}     & 0.800              & 0.733                    & 1                   & 0.733             & 0.733           & 0.800         & 0.800                   \\ \hline
\textbf{Plus-One (PO)}       & 1                  & 1                        & 1                   & 1                 & 1               & 1             & 1                       \\ \hline
\textbf{Regular (R)}        & 1                  & 0.840                    & 0.820               & 0.820             & 1               & 0.520         & 0.820                   \\ \hline
\textbf{Short (S)}          & 0.580              & 0.580                    & 0.400               & 0.400             & 0.520           & 1             & 0.420                   \\ \hline
\textbf{Slow Increasing (SI)} & 0.938              & 0.938                    & 0.938               & 0.938             & 0.938           & 1             & 1                       \\ \hline
\end{tabular}
\end{table}

\clearpage 
\bibliographystyle{naturemag}
\bibliography{bib}